\documentclass[aps,prd,superscriptaddress,12pt,showpacs,notitlepage]{revtex4}
\usepackage{graphics}
\usepackage{amsfonts}
\usepackage{amsmath,amssymb,mathrsfs}
\usepackage[dvips]{graphicx}
\usepackage{subfig}

\newcommand{\be}{\begin{equation}}
\newcommand{\ee}{\end{equation}}
\newcommand{\bea}{\begin{eqnarray}}
\newcommand{\eea}{\end{eqnarray}}
\newcommand{\pa}{\partial}
\newcommand{\bb}{\bibitem}
 
\begin{document}
\title{The gauged BPS baby Skyrme model}
\author{C. Adam}
\affiliation{Departamento de F\'isica de Part\'iculas, Universidad de Santiago de Compostela and Instituto Galego de F\'isica de Altas Enerxias (IGFAE) E-15782 Santiago de Compostela, Spain}
\author{C. Naya}
\affiliation{Departamento de F\'isica de Part\'iculas, Universidad de Santiago de Compostela and Instituto Galego de F\'isica de Altas Enerxias (IGFAE) E-15782 Santiago de Compostela, Spain}
\author{J. Sanchez-Guillen}
\affiliation{Departamento de F\'isica de Part\'iculas, Universidad de Santiago de Compostela and Instituto Galego de F\'isica de Altas Enerxias (IGFAE) E-15782 Santiago de Compostela, Spain}
\author{A. Wereszczynski}
\affiliation{Institute of Physics,  Jagiellonian University,
Reymonta 4, Krak\'{o}w, Poland}


\begin{abstract}
The baby Skyrme model is a well-known nonlinear field theory supporting topological solitons in two space dimensions. In the limit where the term quadratic in derivatives (the "sigma model term") vanishes some additional structure emerges. The resulting ("extreme" or "restricted" or "BPS") baby Skyrme model has exact soliton solutions saturating a BPS bound which exists for this restricted model. Further, the restricted model has infinitely many symmetries and infinitely many conservation laws. Here we consider the gauged version of the restricted baby Skyrme model with gauge group U(1) and the usual Maxwell term for the gauge field. We find that, again, there exists a BPS bound and BPS solutions saturating this bound. We further find that the whole problem is essentially determined by a new kind of superpotential equation. The BPS bound and the corresponding BPS solitons only may exist for potentials such that the superpotential equation has a solution which exists globally, i.e., on the whole target space. We also calculate soliton solutions both exactly and numerically, completely confirming our qualitative analytical results.

\end{abstract}

\maketitle 

\section{Introduction}

It is a notoriously difficult problem to derive the basic physical properties of hadrons and nuclei from the underlying fundamental theory of strong interactions, i.e., quantum chromodynamics (QCD). One important approach to resolve this problem consists in the introduction of low-energy effective field theories, where the fundamental fields of QCD, which are the relevant degrees of freedom at high energies, are replaced by some other degrees of freedom which are supposed to describe strong interaction physics at low energies more practically or more concisely. The Skyrme model \cite{skyrme}
is one well-known proposal for such an effective field theory, where the basic degrees of freedom consist of a triplet of chiral fields (Skyrme fields) which are related to the Goldstone bosons of chiral symmetry breaking in the underlying QCD. The nucleons and nuclei, on the other hand, emerge as collective excitations of the basic fields, i.e., topological soliton solutions (called skyrmions) which the theory supports. Further, the topological degree of the solitons is identified with the baryon number. Quantization of the spin and isospin degrees of freedom for the simplest solitons leads to predictions for the physical properties of  nucleons and light nuclei which are in reasonable agreement with experiment, see, e.g.,  \cite{AdNaWi}, \cite{AdNa}, \cite{BraCar} (for further, more recent results see, e.g., \cite{man1}). These early successes have lead to an extensive numerical study of soliton solutions, both for the standard Skyrme model (consisting of a non-linear sigma model term quadratic in first derivatives and a so-called Skyrme term which is quartic in first derivatives) and for the standard Skyrme model with an additional potential term ("pion mass term"). As a result, solitons in the standard Skyrme model for topological degree up to about 20 are now fairly well understood \cite{BaSu1}, \cite{BaSu2}, \cite{man-sut-book}. 

Although the Skyrme model lagrangian cannot be rigorously derived from QCD, the symmetries and anomalies of QCD provide sufficient information to uniquely determine the coupling of the Skyrme field to electromagnetism \cite{wit1}. Concretely, the gauge field couples both to the topological (baryon) current and to the third component of the isospin current. A first attempt to determine the electromagnetic contributions to the proton and neutron masses within the Skyrme model perturbatively in the electromagnetic field and within the Couloumbic approximation has been done in \cite{DurPak}. Unsurprisingly, the electrostatic energy of the (charged) proton is larger than that of the (electrically neutral) neutron. A correct calculation of the proton-neutron mass difference most likely requires some amount of explicit breaking of the isospin symmetry. 
A mainly numerical analysis of the standard Skyrme model with the electromagnetic field coupled to the third component of the isospin has been performed in \cite{pie-tra}, and some further results have been presented in \cite{radu-tra}. 

To the best of our knowledge, the analysis of the fully nonlinear coupled Skyrme-Maxwell system has not yet been carried to the point where reliable descriptions of the electromagnetic properties of nucleons and light nuclei would be possible, despite the obvious physical interest of such a description, most likely due to the tremendous difficulty of this task. It is at this point where the study of lower-dimensional models becomes useful or even indispensable.
The investigation of lower dimensional  field theories is a well-established
research field, both in order to understand  and solve difficult nonlinear
 problems, and because of the growing appearance of interesting physical
realizations of planar physical systems.  The simplifications which may be achieved in lower dimensions help to
achieve a better visualization of the system and  often allow to avoid unessential technichal
complications, even though one has to remain aware of potential
oversimplifications. One good example
is  the problem of stability and order reduction  (via BPS equations) of the
nonlinear sigma and Skyrme models, which was considered in three dimensions by \cite{BePo}, leading to the discovery of instanton solutions.

More specifically, there exists a 2+1 dimensional version of the Skyrme model which was originally introduced in \cite{old}, \cite{holom}, and further investigated, e.g., in \cite{new}, \cite{other}, \cite{kudr} (for more recent results see, e.g., \cite{karliner} - \cite{foster1}). In this so-called baby Skyrme model, the dimensions both of the base space and of the target space are reduced by one, such that a topological degree characterizing field configurations and topological soliton solutions exist, again, in close analogy to the full Skyrme model. This model has already found some independent physical applications, e.g., in condensed matter systems \cite{qhe}, or in brane cosmology where the solitons of the model induce co-dimension two branes \cite{sawa1}. 
The lagrangians of the Skyrme and baby Skyrme models are very similar, as well. In both models there exists a nonlinear sigma-model type kinetic term and a Skyrme term which is quartic in first derivatives. In addition, the potential term is mandatory in the baby Skyrme model for soliton solutions to exist, as a consequence of the Derrick scaling argument. The baby Skyrme model, too, has a natural coupling to the electromagnetic field. It turns out that the resulting baby Skyrme-Maxwell system in one lower dimension does allow for a nonlinear treatment with a reasonable numerical effort.  Indeed, soliton solutions of the full coupled system have been calculated in \cite{GPS}, and their electric and magnetic properties have been studied, too. Further, it turns out that in the gauged system the Skyrme term is not mandatory because the Maxwell term shows the same scaling behaviour, and the resulting "gauged nonlinear sigma model with a potential" obtained by skipping the Skyrme term has soliton solutions, too, see \cite{schr1}. 
Both the gauged Skyrme model and the gauged nonlinear sigma model with potential have a BPS bound for the energy in terms of the topological degree, but nontrivial soliton solutions, in general, do not saturate this bound. Still, the gauged nonlinear sigma model does have genuine BPS soliton solutions for a specific choice of the potential \cite{schr1} (see Section II). Apparently, the progress in the understanding of these 2+1 dimensional models is in part related to the reduced number of dimensions and has not yet provided us with all the necessary tools for a deeper analysis of the full 3+1 dimensional Skyrme-Maxwell system. So one might wonder whether there exists some further structure which is shared by the baby and the full Skyrme theory and which might bring us closer to this final goal.  It is the purpose of the present article to give an affirmative answer to this question, but before presenting our proposal it is necessary to briefly review some recent results on BPS submodels of the Skyrme and baby Skyrme models.

The field space (or target space) of the baby Skyrme model is given by the two-sphere $S^2$. If we require that the Lagrangian density for this field variable depends on the fields and their first derivatives, is Poincare invariant and at most quadratic in time derivatives - in order to allow for a standard hamiltonian formulation - then the Lagrangian of the baby Skyrme model is the most general possibility. Specifically, the Skyrme term which is quartic in derivatives and is, at the same time, the square of the topological current density, already has the highest possible power of derivatives. The simplest submodel supporting topological solitons is the scale-invariant nonlinear sigma model \cite{BePo} consisting only of the quadratic term, but otherwise both the potential and the quartic Skyrme term are required to maintain stability under Derrick scaling. The quadratic (sigma model) term, on the other hand, is not mandatory from this point of view, and the restricted model containing only the potential and the Skyrme term not only supports solitons but, additionally, has some further structure. It was first considered in \cite{GP} where the infinitely many base space symmetries of the static energy functional were found, and both exact static soliton solutions and exact time-dependent topological Q-ball solutions were calculated. Then, in \cite{restr-bS} it was found that this restricted baby Skyrme model has both a BPS bound and BPS soliton solutions saturating this bound. Further, it was demonstrated that the model has infinitely many target space Noether symmetries with their corresponding conservation laws (it has, in fact the zero curvature representation of generalized integrability \cite{gen-int}). A more geometric interpretation of the BPS bound and the corresponding BPS equations was given in \cite{Sp1}.  

If the same conditions as above are imposed for the $S^3$ (or SU(2)) field space of the Skyrme model, then the most general lagrangian density consists of a potential term, the quadratic nonlinear sigma model term, the quartic Skyrme term and a sextic term which is, at the same time, the square of the topological current density. That is to say, the quartic Skyrme term of the baby Skyrme model has two possible generalizations which both are equally acceptable from the point of view of the Derrick scaling argument. The full generalized Skyrme model with all allowed terms present has been considered and applied to nucleons and nuclei (see e.g. \cite{Jack1}, \cite{Flor1}, \cite{DiYa1}), but a systematic search of higher solitons, analogous to the case of the standard Skyrme model, has not yet been performed to the best of our knowledge. This generalized Skyrme model, again, has a submodel (the "BPS Skyrme model") consisting of a potential and the sextic term, which shows all the additional structure mentioned above \cite{BPS-Sk1}, \cite{BPS-Sk2}. The static energy functional has the volume-preserving diffeomorphism on base space as symmetries, which is interesting from the point of view of nuclear physics (nuclear liquid drop model), because these are precisely the symmetries of an incompressible liquid. Further, the BPS Skyrme model is integrable in the sense of generalized integrability and has an infinite number of target space Noether symmetries and the corresponding conservation laws. Thirdly, the BPS Skyrme model has a BPS bound and soliton solutions saturating this bound. This fact is of special importance because it resolves one of the main problems in the application of the Skyrme model to nuclear physics, namely the too large binding energies of nuclei which the standard Skyrme model predicts. As a consequence of the BPS property, classical soliton solutions in the BPS Skyrme model have zero binding energies, and realistic small values for the binding energies of nuclei may be achieved both by quantum corrections and by small contributions of additional terms in the lagrangian \cite{BoMa}, \footnote{A different way to construct a Skyrme model which has the BPS property via the systematic inclusion of vector mesons has been proposed and investigated in \cite{Sut1}}. 

In view of the existence and of the rich structure of the integrable BPS Skyrme submodels described above, both in 2+1 and in 3+1 dimensions, one rather obvious question is whether these submodels may be gauged and whether the resulting gauged BPS Skyrme models maintain the integrability, the BPS property, and the close relationship between the 2+1 dimensional and the 3+1 dimenional case. It is the purpose of the present article to study the 2+1 dimensional case and to demonstrate that the resulting gauged BPS baby Skyrme model in 2+1 dimensions still is integrable and has both a BPS bound and BPS solitons which, in some cases, may even be calculated exactly. An investigation of the 3+1 dimensional case, following the lines developed in the present article, will be presented elsewhere.

The paper is organized as follows. In Section II, we introduce the gauged baby Skyrme model and briefly recapitulate some known results. Then we restrict to the gauged BPS baby Skyrme model which we want to investigate, derive its Euler--Lagrange equations and discuss the boundary value problem relevant for soliton solutions. In Section III we demonstrate that integrability still holds for the gauged model. In Section IV we derive the BPS property of the gauged BPS baby Skyrme model. Concretely, we derive the BPS bound in Section IV.A, whereas in Section IV.B  we demonstrate that the BPS equations imply the static Euler--Lagrange equations.   
Section IV.C is dedicated to the issue of BPS soliton solutions. In Section V we present the numerical calculations of soliton solutions for several potentials and coupling constants. The soliton energies saturate the BPS bound in all cases. In Section VI we present exact BPS soliton solutions for some specific choices of the potential. Finally, Section VII contains our conclusions. 
We decided to give a rather detailed presentation of the model and its properties in this article, because some of the methods used and some of the results presented are quite different from known ones (e.g. the BPS bound is markedly different from all BPS bounds known to us), and a detailed understanding of the new features of this model and the new analytical methods will be indispensable both for the study of the analogous case in 3+1 dimensions and for its use in other contexts (e.g., to understand whether the new type of BPS bound presented here may be employed for other field theories). 

\section{The gauged model}
The degrees of freedom of the ungauged baby Skyrme model are described by a three-component vector of scalar fields $\vec \phi = (\phi_1 ,\phi_2 ,\phi_3)$ with unit length, ${\vec{\phi}}^2 \equiv \phi_1^2 + \phi_2^2 + \phi_3^2 =1$. That is to say, its target space manifold is a two-sphere $S^2$. The Lagrangian of the (ungauged) standard baby Skyrme model consists of three terms, namely the nonlinear sigma model term $L_2$ quadratic in derivatives, the Skyrme term $L_4$ (which is quartic in first derivatives), and the potential term with lagrangian density ${\cal L}_0 = -\lambda_0 V(\vec\phi )$. Here, the potential $V(\vec \phi)$ always is a real, nonnegative function of its arguments. Both $L_2$ and $L_4$ are invariant under the full group of target space rotations SO(3) (i.e., rotations acting on $\vec \phi$), whereas the potential necessarily breaks part of this symmetry. Throughout this paper, we shall be concerned only with potentials which preserve a U(1) subgroup of this target space symmetry and which, further, have a unique vacuum (i.e., there is no spontaneous symmetry breaking). Concretely, we assume that
\be
V = V(\vec n\cdot \vec \phi )  \; ,\quad V(1)=0,
\ee
which implies that $V$ is invariant w.r.t. rotations about the axis $\vec n$ and that $\vec \phi = \vec n =$ const. is the (unique) vacuum configuration. In concrete calculations we shall always assume
\be
\vec n = (0,0,1) \quad \Rightarrow \quad V=V(\phi_3) 
\ee
such that the vacuum value is $\phi_3 =\phi_{3,vac}=1$, but in more general expressions it will be useful to maintain the general vacuum vector $\vec n$. 
Finite energy field configurations $\vec \phi (t,\vec x)$ have to approach the vacuum configuration in the limit of large $|\vec x|$ independent of the direction of $\vec x$, $\lim_{|\vec x |\to \infty} \vec \phi (t,\vec x)=\vec n$ for all times. This allows for the one-point compactification of the physical base space $\mathbb{R}^2$ which makes it topologically equivalent to the two-sphere $S^2$. As a consequence, finite energy field configurations may be interpreted as maps $S^2 \to S^2$ which are characterized by an integer winding number (or topological degree)
\be
{\rm deg}[\vec \phi] = \frac{1}{4\pi} \int d^2 x \vec \phi \cdot \pa_1 \vec \phi \times \pa_2 \vec \phi = k \; ,\quad k \in \mathbb{Z}.
\ee
Due to its integer-valuedness this winding number is obviously conserved in time.

\subsection{The gauged baby Skyrme model}
The presence of the unbroken subgroup U(1) opens the possibility to gauge this subgroup by a U(1) gauge field whose dynamics is governed by the usual Maxwell term. The right coupling of the gauge field to the Skyrme field $\vec \phi$ is achieved by replacing ordinary partial derivatives with the covariant derivatives
\be
D_\alpha \vec \phi \equiv \partial_\alpha \vec \phi + A_\alpha \vec n \times \vec \phi
\ee
as may be checked easily \cite{schr1}, \cite{GPS}. 
The resulting Lagrangian density (i.e., the gauged baby Skyrme model) is
\be
{\cal L}= {\cal L}_2 + {\cal L}_4 + {\cal L}_0 + {\cal L}_M
\ee
where ${\cal L}_2$ is the gauged sigma model term
\be
{\cal L}_2 = \frac{\lambda_2}{2}(D_\alpha \vec \phi )^2 ,
\ee
${\cal L}_4$ is the Skyrme term
\be
{\cal L}_4 = -\frac{\lambda_4}{4}(D_\alpha \vec \phi \times D_\beta \vec \phi )^2 ,
\ee
and ${\cal L}_0$ is the potential
\be
{\cal L}_0 = -\lambda_0 V(\vec n \cdot \vec \phi ).
\ee
Finally, ${\cal L}_M$ is the usual Maxwell term
\be
{\cal L}_M = -\frac{\lambda_M}{4}F_{\alpha\beta}^2 \; , \quad F_{\alpha\beta} = \pa_\alpha A_\beta - \pa_\beta A_\alpha .
\ee
Here, all fields are defined on 2+1 dimensional Minkowski space, and we use the metric with signature $(+,-,-)$. Further, the $\lambda_k$, $k=0,2,4,M$ are non-negative dimensionful coupling constants. Following \cite{GPS}, we now introduce new coupling constants by extracting a common energy scale $E_0$ from all terms in the Lagrangian. The resulting Lagrangian is
\be
L=E_0 \int d^2 x \left( \frac{\nu^2}{2} (D_\alpha \vec \phi )^2 - \frac{\lambda^2}{4} (D_\alpha \vec \phi \times D_\beta \vec \phi )^2 
-\mu^2 V(\vec n\cdot \vec \phi ) - \frac{1}{4g^2} F_{\alpha\beta}^2 \right) .
\ee
Here, $E_0$ has the dimension of energy and sets the energy scale of our model. In concrete calculations we shall always express energies in units of $E_0$, which is equivalent to setting $E_0=1$. $\nu$ is a dimensionless constant which takes the value $\nu =1$ in the original gauged baby Skyrme model of \cite{GPS} (for a nonzero $\nu$, $\nu =1$ can always be achieved by an appropiate choice of the energy scale $E_0$), whereas we shall choose $\nu =0$ for the case of our restricted (or BPS) gauged baby Skyrme model. Further, $\lambda$ has the dimension of length, whereas $\mu$ and $g$ have the dimensions of inverse length. $g$ is, in fact, the electromagnetic coupling constant. 

In this paper we are mainly interested in static (soliton) solutions, so let us restrict to the static energy functional in what follows. Further, it has been demonstrated already in \cite{GPS} that static finite energy solutions must have zero electric field, at least for the spherically symmetric ansatz considered there. We shall assume this in our paper, i.e., we choose for the static gauge field
\be
A_\alpha (\vec x) = (0,A_1 (\vec x) ,A_2 (\vec x)) \; ,\quad E_1 =E_2 =0 \; ,\quad B(\vec x) = \pa_1 A_2 - \pa_2 A_1
\ee
where $B$ is the magnetic field in 2+1 dimensions and the electric field $E_i$ is zero. Then the static energy functional which defines our static variational problem is
\be
 E\equiv E^V (\nu,\lambda,\mu,g)=\frac{1}{2}E_0 \int d^2 x \left( \nu^2 (D_i \vec \phi )^2 + \lambda^2 (D_1 \vec \phi \times D_2 \vec \phi )^2 +
2\mu^2 V(\vec n\cdot \vec \phi ) + \frac{1}{g^2} B^2 \right) .
\ee

\subsection{Some known energy bounds}
Before starting our calculations, it is useful to review some known results. It is well-known that the ungauged pure nonlinear sigma model with energy functional
\be
E_u (\nu, \lambda =0,\mu =0) = \frac{1}{2} E_0 \nu^2 \int d^2 x  (\pa_i \vec \phi )^2
\ee
(here the subindex $u$ stands for "ungauged", i.e., zero gauge field $A_\alpha =0$) has a topological lower energy bound for static configurations
\cite{BePo}
\be
E_u (\nu, \lambda =0,\mu =0) \ge 4\pi E_0 \nu^2 |{\rm deg} [\vec \phi ]| .
\ee
Further, the pure nonlinear sigma model has the meromorphic functions as static solutions saturating this bound (i.e., BPS solutions; in fact, all static finite energy solutions are BPS solutions). The same bound remains true for the full ungauged baby Skyrme model,
\be
E_u^V (\nu ,\lambda ,\mu )\ge 4\pi E_0 \nu^2 |{\rm deg} [\vec \phi ]|
\ee
because of the obvious inequality $E_u^V (\nu ,\lambda ,\mu ) \ge E_u (\nu ,\lambda =0,\mu =0 )$. Nontrivial solutions of the full baby Skyrme model do not saturate this bound. There exists an additional, more stringent bound 
for the full, ungauged baby Skyrme model which derives from the fact that there exists a second, independent bound for the baby Skyrme model without the sigma model term, i.e., for the BPS baby Skyrme model $E_u^V (\nu=0,\lambda ,\mu)$ (see Section IV). Again, nontrivial soliton solutions of the full baby Skyrme model in general do not saturate this bound. 

For the gauged baby Skyrme model, there exist some known BPS bounds, as well. First of all, for the gauged model without the quartic Skyrme term (i.e., for $\lambda =0$) there exists a specific choice of potential and coupling constants such that the resulting "gauged nonlinear sigma model with potential" has both a BPS bound and BPS solutions saturating this bound \cite{schr1}. The specific potential is
\be \label{V-s}
V_s (\vec n \cdot \vec \phi )= \frac{1}{2} \left( 1-\vec n \cdot \vec \phi \right)^2
\ee 
and the coupling constants must be chosen equal, $\mu=g$ (in addition to $\lambda =0$; further we set $\nu =1$ without loss of generality). The resulting energy is
\bea
E^{V_s} (\nu =1, \lambda =0,\mu ,g=\mu)  &=& E^{V_s} (\nu =1, \lambda =0,\mu =1,g=1) \nonumber \\
&=&
\frac{1}{2}E_0 \int d^2 x \left(  (D_i \vec \phi )^2   +
2 V_s(\vec n\cdot \vec \phi ) +  B^2 \right)
\eea
where we transformed $\mu$ and $g=\mu$ into $\mu = g =1$ by a variable transformation $\vec x \to \mu \vec x$ to dimensionless new variables $\vec x$. It was proved in \cite{schr1} that this model obeys the same BPS bound like the ungauged pure nonlinear sigma model,
\be
E^{V_s} (\nu =1, \lambda =0,\mu =1,g=1) \ge 4\pi E_0 |{\rm deg} [\vec \phi] |
\ee
and that there exist BPS solutions saturating the bound for winding number $|{\rm deg}[\vec \phi ]|>1$. For $|{\rm deg}[\vec \phi ]|=1$, on the other hand, finite energy solutions do not exist. 

The above BPS bound for the specific potential $V_s$ may be used, under certain conditions, to derive energy bounds for other potentials. To show it, we first rewrite the energy expression like
\be
E^V (\nu =1,\lambda =0, \mu ,g)=E_0 \int d^2 x \left(  (D_i \vec \phi )^2   +
2 \frac{\mu^2}{g^2} V(\vec n\cdot \vec \phi ) +  B^2 \right)
\ee
where we transformed again to new dimensionless space coordinates $\vec x \to g \vec x$. If $\frac{\mu^2}{g^2}V$ may be bound by multiples of $V_s$, i.e., $\frac{\mu^2}{g^2}V \ge c_V V_s$, then the above energy may be bound either by $4\pi E_0 |{\rm deg}[\vec \phi ]|$ (if $c_V \ge 1$) or by $4\pi E_0 c_V |{\rm deg}[\vec \phi ]|$ (if $c_V < 1$). Let us choose the so-called "old baby Skyrme potential" 
\be \label{V-o}
V_o (\vec n \cdot \vec \phi )= 1-\vec n\cdot \vec \phi
\ee
as an example. As a consequence of the inequality
\be
V_o \equiv 1-\vec n \cdot \vec \phi \ge \frac{1}{2} \left( 1-\vec n\cdot \vec \phi \right)^2 \equiv V_s
\ee
we find the following energy bound
\bea
\frac{\mu}{g} \ge 1 : && E^{V_o} (\nu =1,\lambda =0, \mu ,g) \ge 4\pi E_0 |{\rm deg}[\vec \phi ]| \nonumber \\
\frac{\mu}{g} < 1 : && E^{V_o} (\nu =1,\lambda =0, \mu ,g) \ge 4\pi E_0 \frac{\mu^2}{g^2} |{\rm deg}[\vec \phi ]| .
\eea
In these models, however, soliton solutions will in general not saturate the bound, i.e., they are not of the BPS type.

These bounds continue to hold for the full gauged baby Skyrme model, i.e., for $\lambda \not= 0$, as a consequence of the obvious inequality
$E^V (\nu =1,\lambda , \mu ,g) \ge E^V (\nu =1,\lambda =0, \mu ,g)$. For the old baby Skyrme model, e.g., we have (see \cite{GPS})
\bea
\frac{\mu}{g} \ge 1 : && E^{V_o} (\nu =1,\lambda , \mu ,g) \ge 4\pi E_0 |{\rm deg}[\vec \phi ]| \nonumber \\
\frac{\mu}{g} < 1 : && E^{V_o} (\nu =1,\lambda , \mu ,g) \ge 4\pi E_0 \frac{\mu^2}{g^2} |{\rm deg}[\vec \phi ]| .
\eea
Again, these bounds are not saturated for nontrivial soliton solutions.

\subsection{The gauged BPS model}
Now we shall restrict to the case we finally want to discuss in detail, i.e., to the case without the sigma model term, $\nu =0$, with lagrangian density
\begin{equation}\label{Lag}
 {\cal L} = -\frac{\lambda^2}{4} \left( D_\alpha \vec \phi \times D_\beta  \vec \phi \right)^2 - \mu^2 V ( \vec n \cdot \vec \phi)
 - \frac{1}{4 g^2} F^2_{\alpha \beta}.
\end{equation}
The Euler--Lagrange (EL) equations are derived by varying w.r.t. the fields and their derivatives, as usual. For the Skyrme field $\vec \phi$ there exists a minor subtlety related to the constraint $\vec \phi^2 =1$. This constraint should in principle be implemented by adding a Lagrange multiplyer term $\lambda (x)(1-\vec \phi^2)$ to the lagrangian density (we suppressed this term for the sake of brevity). As a consequence, the component of the EL equation in the direction of $\vec \phi$ is not dynamical but serves, instead, just to determine the Lagrange multiplier $\lambda (x)$. We may project on the dynamical part of the EL equation by multiplying the full EL equation with $\vec \phi \times$, which leads to
\begin{equation}
 D_\alpha \vec{\mathcal{K}}^\alpha = -\mu^2 \vec n \times \vec \phi \; V' .
\end{equation}
Further, the inhomogeneous Maxwell equation is
\begin{equation}
 \partial_\alpha F^{\alpha \beta} = g^2 \vec n \cdot \vec{\mathcal{K}}^\beta,
\end{equation}
where
\begin{equation}
\vec{\mathcal{K}}^\alpha = \lambda^2 D_\beta \vec \phi \left[ \vec \phi \cdot (D^\alpha \vec \phi \times D^\beta \vec \phi) \right].
\end{equation}
The corresponding energy functional is
\begin{equation}
E=\frac{1}{2} \int d^2 x \left( \frac{\lambda^2}{2} \left( D_{0} \vec{\phi} \times D_{i} \vec{\phi} \right)^2 + \frac{1}{g^2} E_i^2 + \lambda^2 \left( D_{1} \vec{\phi} \times D_{2} \vec{\phi} \right)^2 +2\mu^2V+\frac{1}{g^2}B^2\right) .
\end{equation}
Next we assume $\vec n = (0,0,1)$ and the standard static ansatz
\begin{equation} \label{rad-ans}
\vec{\phi} (r,\phi)  = \left( 
\begin{array}{c}
\sin f(r) \cos n\phi \\
\sin f(r) \sin n\phi \\
\cos f(r)
\end{array}
\right), \;\;\;\; A_0=A_r=0, \;\;\; A_\phi=na(r) 
\end{equation}
then the electric field vanishes identically and $B=\frac{na'(r)}{r}$. Further, this field configuration has winding number ${\rm deg}[\vec \phi] =n$ provided that $f$ obeys the appropriate boundary conditions, see below. The field equations can be reduced to 
\bea
\frac{1}{r^2} f'' (1+a)^2 \sin^2 f + \frac{f'}{r} \left[ \left( 2a'-\frac{1+a}{r} \right) \frac{1+a}{r} \sin^2 f + \frac{f'}{r} (1+a)^2 \sin f \cos f \right]  && \nonumber \\
+ \; \frac{\mu^2}{n^2\lambda^2} \sin f \; V'&=&0 
\eea
 \begin{equation}
 a''-\frac{a'}{r} = \lambda^2 g^2 (1+a) \sin^2 f f'^2 
\end{equation} 
where now $V=V(\phi_3) = V(\cos f)$ and $V' = V_{\phi_3}$. 
Further, we introduce the new variable
\begin{equation}
y=\frac{r^2}{2}
\end{equation}
to arrive at the following system of autonomous second order equations 
\begin{equation}
\sin f \left\{ \partial_y \left[ f_y (1+a)^2\sin f\right] + \frac{\mu^2}{n^2 \lambda^2} V' \right\}=0
\end{equation} 
\begin{equation}
a_{yy}=\lambda^2g^2(1+a)\sin^2 f f_y^2 .
\end{equation} 
With the help of the function
\begin{equation}
\phi_3 = \cos f \equiv 1-2h \; \Rightarrow \; h =\frac{1}{2} (1-\cos f), \;\;\; h_y = \frac{1}{2} \sin f f_y
\end{equation} 
this may be further simplified to
\begin{equation} \label{h-eq}
\sin f \left\{ \partial_y \left[ h_y (1+a)^2 \right] - \frac{\mu^2}{4n^2 \lambda^2} V_h \right\}=0
\end{equation} 
\begin{equation} \label{a-eq}
a_{yy}=\lambda^2g^2(1+a) 4 h_y^2 
\end{equation} 
where now $V=V(h)$ and $V_h =-2V'$. This system of two second order equations has, in general, four integration constants, i.e., a four dimensional solution manifold. We shall see in a moment that, on the other hand, a soliton solution imposes four boundary conditions, therefore generically we expect at most one soliton solution for this spherically symmetric ansatz and for a given $n$ and a given potential. At $y=0$ (i.e., $r=0$), a soliton solution has to obey the two conditions
\be \label{sol-bound-zero}
h(0)=1 \quad \Leftrightarrow \quad f(0)=\pi \; , \quad a(0)=0.
\ee 
The precise form of the further boundary conditions depends on the way the fields approach their vacuum values. If the soliton
takes its vacuum value already at a finite radius $r=r_0$ (i.e., $y=y_0$) and remains in the vacuum for $y \ge y_0$ it is said to be of the compacton type \cite{GP}, \cite{arodz}, \cite{k-comp}, \cite{comp-bS}, \cite{restr-bS}. 
 In this case, the additional boundary conditions are
\be \label{sol-bound-comp}
a_y(y=y_0) =0 \; , \quad h(y_0) = h_y(y_0)=0 .
\ee
Apparently these are three more conditions, but we introduced the additional free constant $y_0$ (the compacton radius), so these correspond to two more conditions, and the total number of boundary conditions is four.  
Alternatively, if the field $h$ approaches the vacuum exponentially (like $\exp -c y$) or if the approach to the vacuum is power-like (like $y^{-c}$) (here $c$ is some positive real constant) then the fields reach their vacuum values in the limit $y\to \infty$, so the boundary conditions for a soliton are
\be \label{sol-bound-inf}
\lim_{y\to \infty} h(y)=0 \; , \quad \lim_{y\to \infty} a_y(y)=0
\ee
whereas $\lim_{y\to \infty} h_y (y)=0$ is not an independent condition but, rather, a consequence of the asymptotic form of the above field equations. Again, the total number of boundary conditions is four.   

We remark that in the ungauged model there is a simple relation between the potential $V(\phi_3)$ and the approach to the vacuum, independent of the coupling constants. Indeed, if $V$ is less than of a fourth power in small flucuations about the vacuum, then a possible soliton solution of the ungauged model must be of the compacton type, i.e., it takes its vacuum value already at a finite radius $r=r_0$ (i.e., $y=y_0$) and remains in the vacuum for $y \ge y_0$, \cite{restr-bS}. Further, if the potential is exactly quartic in small fluctuations, then the approach of a soliton to its vacuum value is exponential, whereas if $V$ goes to zero with a higher than fourth power then a soliton of the ungauged model approaches the vacuum in 
a power-like way.
Observe that $h$ itself is already quadratic in small fluctuations because $h=(1/2)(1-\phi_3) = (1/2)(1-\sqrt{1-\phi_1^2 - \phi_2^2}) \sim (1/4)(\phi_1^2 + \phi_2^2)$. Therefore, a potential which is "quartic in small fluctuations" means a potential which is "quadratic in $h$" for small $h$. 
 The "old" baby Skyrme potential (\ref{V-o}), e.g., is quadratic in small fluctuations, whereas the potential (\ref{V-s}) is quartic. 
At this point the obvious question shows up whether
the simple relation between the potential and the approach to the vacuum continues to hold in the gauged model, or whether the approach to the vacuum will depend both on the potential and on the values of the coupling constants. For the specific case of the so-called old baby Skyrme potential it is easy to see that the potential alone is sufficient to determine the asymptotics, see next subsection. For the general case it is more difficult to answer this question, but we will find strong theoretical indications that still the potential alone is sufficient to determine the asymptotic behaviour, see Section VI.

\subsection{The old baby Skyrme potential}
Although we will consider different potentials, most of our numerical calculations will be done for the old baby Skyrme potential (\ref{V-o}). One reason is that this is the case considered in \cite{GPS}.  A further reason is related to the fact that for the old baby Skyrme potential the system of second order equations (\ref{h-eq}), (\ref{a-eq}) has a simple first integral. Indeed, for the old baby Skyrme potential these two equations read
\begin{equation}
\sin f \left\{ \partial_y \left[ h_y (1+a)^2 \right] - \frac{\mu^2}{2n^2 \lambda^2} \right\}=0
\end{equation} 
\begin{equation}
a_{yy}=\lambda^2g^2(1+a) 4 h_y^2 \label{mag} .
\end{equation} 
The first equation is solved by 
\begin{equation}
\sin f =0 \;\; \Rightarrow \;\; f=0 \;\; \Rightarrow \;\; h=0
\end{equation} 
or by setting the expression in brackets equal to zero, which can be easily integrated to give
\begin{equation}
h_y(1+a)^2 = \frac{\mu^2}{2n^2 \lambda^2} (y-y_0). \label{profile}
\end{equation} 
It already follows from this first integral that $y_0$ is the compacton radius and that the nontrivial solution of the above equation for $y\le y_0$ must be joined with the vacuum solution $h=0$ for $y>y_0$ if we want to avoid a solution which grows indefinitely for large $y$. Formally, this first integral
may be further integrated to give 
\begin{equation}
h(y) = \int dy  \frac{\mu^2}{2n^2 \lambda^2} \frac{(y-y_0)}{(1+a(y))^2} +\alpha_0.
\end{equation} 
The magnetic function $a$ can be obtained from eq. (\ref{mag}), where the baby skyrmion profile function is expressed in terms of $a$ by (\ref{profile}). Then, we get
\begin{equation}
(1+a)^3 a_{yy}=\lambda^2g^2 \left(  \frac{\mu^2}{n^2 \lambda^2}\right)^2  (y-y_0)^2
\end{equation} 
or 
\begin{equation}
(1+a)^3 a_{yy}=0
\end{equation} 
outside of the compact baby skyrmion.  It gives a trivial solution outside the compacton 
\begin{equation}
a(r \geq r_0) =a_R=const. 
\end{equation}
This implies that also the magnetic field is confined inside the compact baby skyrmion. Hence, we still have genuinely compact objects which interact only via a contact interactions. Obviously we may also construct multisoliton configurations exactly as in the standard BPS baby Skyrme model.
The equation for $a$ can be simplified by introducing $b(y)=1+a(y)$ and $\beta =\lambda^2g^2 \left(  \frac{\mu^2}{n^2 \lambda^2}\right)^2$. Further $z=\beta^{1/4}(y-y_0)$, then
\begin{equation}
b^3b_{zz}=z^2.
\end{equation}
Unfortunately, we were not able to find analytic solutions to this equation, so some numerics is still required. Of course, a solution of the magnetic equation should be inserted into the first order equation (\ref{profile}), where we also have to implement the boundary conditions for a compacton solution. Numerical solutions will be calculated and described in detail in section V. 

We remark that the possibility to partially integrate the static field equations already points towards the possible existence of a BPS bound and BPS equations for the static system. We shall see later, in section IV, that this is indeed the case, i.e., there exists a BPS bound, and soliton solutions are, in fact, solutions to the related BPS equations. The existence of such BPS solutions will, however, be related to some nontrivial conditions for the potential which not all potentials satisfy. As a consequence, there exist potentials such that the ungauged models has BPS soliton solutions, whereas the gauged model with the same potential does not support soliton solutions, even for arbitrarily small electromagnetic coupling constant $g$.      

\section{Integrability, symmetries and conservation laws}
The ungauged model
\be
{\cal L} = -\frac{\lambda^2}{4} \left( \partial_\alpha \vec \phi \times \partial_\beta  \vec \phi \right)^2 - \mu^2 V ( \vec n \cdot \vec \phi)
\ee
has an infinite number of target space symmetries with their corresponding Noether currents and conservation laws \cite{restr-bS}. The model possesses, in fact, the zero curvature representation of generalized integrability \cite{gen-int}. It is not difficult to understand the geometric origin of these symmetries. The quartic Skyrme term alone is invariant under the full group of area-preserving diffeomorphisms of the target space $S^2$, because it is just the square of the pullback of the corresponding area two-form \cite{restr-bS}. The potential is only invariant under the subgroup which does not change $\vec n\cdot \vec \phi =\phi_3$, i.e., under the abelian subgroup \cite{ab-dif}
\bea 
\phi_1 & \to & \phi_1 ' = \cos f(\phi_3) \phi_1 - \sin f(\phi_3) \phi_2  \nonumber \\
\phi_2 & \to & \phi_2 ' = \sin f(\phi_3) \phi_1 + \cos f(\phi_3) \phi_2 \nonumber \\
\phi_3 & \to & \phi_3 ' = \phi_3 \label{phase-tr}
\eea
where $f(\phi_3)$ is an arbitrary function of its argument, which makes the symmetry group still infinite dimensional. It turns out that the gauged model maintains exactly the same abelian symmetry group and the corresponding conserved currents. 
\\
Remark: It might appear that in the gauged model the above transformations are just a subset of the full group of gauge transfomations and, therefore, should not provide conservation laws, because it is a well-known fact that gauge transformations do not give rise to nontrivial conservation laws. This is, however, not true. The important point is that the gauged model is separately invariant under the above transformations (\ref{phase-tr}) and under the transformations 
\be \label{phase-tr-a}
A_\alpha \to A_\alpha '  =  A_\alpha - \partial_\alpha g(\phi_3)
\ee
whereas only the joint transformations (\ref{phase-tr}) and (\ref{phase-tr-a}) with $f=g$ are gauge transformations. 
It follows that the transformations (\ref{phase-tr}) are genuine symmetry transformations which give rise to conserved Noether currents and conserved charges. As a consequence of the above argument, the symmetry transformations (\ref{phase-tr-a}) provide exactly (minus) the same conserved charges. 

For an explicit calculation of the conserved currents using the methods of generalized integrability it is useful to switch to the $CP^1$ formulation
of the gauged BPS baby Skyrme model. The Lagrangian density reads
\begin{equation}
\mathcal{L}=\lambda^2 \frac{1}{(1+|u|^2)^4} \left( D_{\alpha} u D_{\beta} u^* -D_{\alpha} u^* D_{\beta} u \right)^2 - \mu^2 V(uu^*) +\frac{1}{4g^2} F_{\alpha \beta}^2
\end{equation}
where the complex scalar field $u$ is related to the unit vector $\vec \phi$ via stereographic projection,
\be
\vec{\phi}=\frac{1}{1+|u|^2} \left( u+\bar{u}, -i ( u-\bar{u}),
|u|^2-1 \right) 
\ee
and the covariant derivatives are
\begin{equation}
D_\alpha u =u_\alpha -ieA_\alpha u, \;\;\; D_\alpha u^* =u_\alpha^* +ieA_\alpha u^* .
\end{equation}
The currents resulting from generalized integrability read
\begin{equation}
j_{\alpha} = iG' (u^*\pi^*_{\alpha}-u\pi_{\alpha})
\end{equation}
where $G=G(uu^*)$ is an arbitrary function of its argument and $\pi_{\alpha}$ is the canonical momentum
\begin{equation}
\pi^*_{\alpha}=\frac{\partial \mathcal{L}}{\partial u^{*  \alpha}} = \frac{4\lambda^2}{(1+|u|^2)^4} (D_\beta u D_\alpha u^* - D_\beta u^* D_\alpha  u) D^\beta u
\end{equation}
\begin{equation}
\pi_{\alpha}=\frac{\partial \mathcal{L}}{\partial u^{\alpha}} = -\frac{4\lambda^2}{(1+|u|^2)^4} (D_\beta u D_\alpha u^* - D_\beta u^* D_\alpha u) D^\beta u^* .
\end{equation}
The field equations are 
\begin{equation}
D_{\alpha} \pi^{\alpha}=\lambda^2 \frac{4u^*}{(1+|u|^2)^5} \left( D_{\alpha} u D_{\beta} u^* -D_{\alpha} u^* D_{\beta} u \right)^2+\mu^2 V'u^*,
\end{equation}
\begin{equation}
D_{\alpha} \pi^{* \alpha} = \lambda^2 \frac{4u}{(1+|u|^2)^5} \left( D_{\alpha} u D_{\beta} u^* -D_{\alpha} u^* D_{\beta} u \right)^2+ \mu^2 V'u .
\end{equation}
Then,
\begin{equation}
\partial_{\alpha} j^\alpha = iG'(u^*_\alpha \pi^{*\alpha} - u_\alpha \pi^\alpha +u^*\partial_\alpha \pi^{* \alpha} -u \partial_{\alpha} \pi^\alpha ) + iG'' (uu^*_{\alpha} + u^* u_\alpha ) (u^*\pi^*_{\alpha}-u\pi_{\alpha}) .
\end{equation}
The first parenthesis is 
\begin{equation}
...=u^*_\alpha \pi^{*\alpha} - u_\alpha \pi^\alpha +ieA_\alpha u^*\pi^{*\alpha} + ieA_\alpha u\pi^\alpha - \mu^2 V'uu^*+ \mu^2 V' uu^*+
\end{equation}
\begin{equation}
+\lambda^2 \frac{4uu^*}{(1+|u|^2)^5} \left( D_{\alpha} u D_{\beta} u^* -D_{\alpha} u^* D_{\beta} u \right)^2 -\lambda^2 \frac{4uu^*}{(1+|u|^2)^5} \left( D_{\alpha} u D_{\beta} u^* -D_{\alpha} u^* D_{\beta} u \right)^2
\end{equation}
\begin{equation}
= D_\alpha u^* \pi^{* \alpha} - D_\alpha u \pi^\alpha =  \frac{4\lambda^2}{(1+|u|^2)^4} (D_\beta u D_\alpha u^* - D_\beta u^* D_\alpha u)(D^\beta u D^\alpha u^* + D^\alpha u D^\beta u^*) =0
\end{equation}
as we contract an antisymmetric tensor with a symmetric one. Further, the second term leads to
\begin{equation}
...=iG''(uu^*_{\alpha} + u^* u_\alpha )\frac{4\lambda^2}{(1+|u|^2)^4} (D^\beta u D^\alpha u^* - D^\beta u^* D^\alpha u) (u^* D_\beta u +uD_\beta u^*)
\end{equation}
\begin{equation}
=iG''\frac{4\lambda^2}{(1+|u|^2)^4} (D^\beta u D^\alpha u^* - D^\beta u^* D^\alpha u) (uu^*_{\alpha} + u^* u_\alpha ) (uu^*_{\beta} + u^* u_\beta )=0.
\end{equation}
Therefore, the current is conserved for arbitrary $G(uu^*)$, and there exists an infinite number of conservation laws, as announced. 

The static energy functional of the ungauged model has the group of area-preserving diffeomorphisms on base space as an additional group of (non-Noether) symmetries. 
Let us now demonstrate that this symmetry, too, is maintained in the gauged model. First of all, the Skyrme term of the gauged model is obtained from the Skyrme term of the ungauged model by simply replacing the partial derivatives $\partial_j$ by covariant derivatives $D_j$. Further, $\partial_j$ and $D_j$ have the same behaviour under coordinate transformations (both are covectors), so the transformation of the Skyrme term under general coordinate transformations (and, therefore, also under the subgroup of are-preserving diffeomorphisms) is the same for the gauged and the ungauged case. For the potential, obviously nothing changes. The potential itself is a scalar and is, therefore, invariant under general coordinate transformations. The invariance of the static energy functional under area-preserving diffeomorphisms follows, therefore, from the same invariance of the area two-form $d^2 x$. Finally, the gauged model contains the Maxwell term. But for a static, purely magnetic configuration $A_\mu = (0,A_1 (\vec x),A_2 (\vec x))$, the Maxwell term is proportional to $B^2$ where $B$ is the magnetic field and transforms like a pseudoscalar under coordinate transformations on the two-dimensional base space of static configurations. It is, therefore, invariant under orientation-preserving diffeomorphisms and, consequently, under area-preserving diffeomorphisms. We conclude that the static energy functional of the gauged model, like the ungauged one,  is invariant under area-preserving base space diffeomorphisms and shows the same degeneracy among field configurations with different shapes but the same area.

\section{The BPS bound}
The BPS bound of the ungauged model was first found in \cite{ward} (as a contribution to an improved bound for the full baby Skyrme model), whereas the BPS nature of the restricted model was proved in \cite{restr-bS} and in \cite{Sp1}. Here we follow the geometric discussion of \cite{Sp1}. 
In a first step, we want to briefly recall the BPS bound of the ungauged model, because we will need this result later on.
The static energy functional of the ungauged model is
\begin{eqnarray}
E&=&\frac{1}{2} E_0 \int d^2 x \left[ \lambda^2 \left( \partial_{1} \vec{\phi} \times \partial_{2} \vec{\phi} \right)^2 +2 \mu^2 V(\phi \cdot \vec n) \right] \\
&=& \frac{1}{2} E_0 \int d^2 x \left[ \lambda^2 q^2 +2 \mu^2 V(\phi_3)  \right]
\end{eqnarray}
where
\begin{equation}
q \equiv \vec \phi \cdot \partial_1 \vec \phi \times \partial_2 \vec \phi
\end{equation}
(i.e., $q$ is $4\pi$ times the topological charge density). This energy functional leads to the bound
\be
E=\frac{1}{2}E_0 \int d^2 x \left( \lambda q \pm \mu \sqrt{2V}\right)^2 \mp E_0 \lambda \mu \int d^2 x q \sqrt{2V} \ge 
\mp E_0 \lambda \mu \int d^2 x q \sqrt{2V}
\ee
with equality if the BPS equation
\be \label{BPSeq-ung}
\lambda q \pm \mu \sqrt{2V} =0
\ee
is satisfied. It remains to show that the bound is, in fact, topological (i.e. does not depend on the field configuration $\vec \phi$). This follows from the following fact. As said already, the Skyrme field $\vec \phi$ defines a map ${\bf \Phi}$ from the one-point compactified base space $\mathbb{R}^2$ to the target space $S^2$, ${\bf \Phi} : \mathbb{R}^2 \to S^2$.
Further, the two-form $d^2 x q $ is just the pullback under this map of the area two-form $\Omega$ on the target space $S^2$, i.e., 
\be
d^2 xq= {\bf \Phi}^* (\Omega) .
\ee 
It follows that $\int d^2 x q \sqrt{2V}$ is just $4\pi$ (i.e., the area of the two-sphere) times the average value of $\sqrt{2V}$ on target space times the times $\vec \phi$ covers the target space while $\vec x$ covers the base space once (i.e., the topological degree or winding number), that is
\be
\int d^2 x q \sqrt{2V(\phi_3)} = 4\pi {\rm deg}[\vec \phi] \langle \sqrt{2V} \rangle_{S^2} .
\ee 
Obviously, this expression does not depend on the field configuration $\vec \phi (\vec x)$. For a given potential $V(\phi_3)$, the average value $\langle \sqrt{2V} \rangle_{S^2}$ takes a fixed, given value, so this expression only depends on the winding number ${\rm deg}[\vec \phi]$, i.e., it is a topological quantity.
The corresponding energy bound reads
\be
E\ge 4\pi \lambda\mu E_0 |{\rm deg}[\vec \phi] \langle \sqrt{2V}\rangle_{S^2}|.
\ee 
Further, there exist BPS soliton solutions which satisfy the BPS equation (\ref{BPSeq-ung}) and, therefore, saturate this BPS bound, see \cite{restr-bS}, \cite{Sp1}.

\subsection{The BPS bound}
The static energy functional of the gauged model is
\begin{eqnarray}
E&=&\frac{1}{2} E_0 \int d^2 x \left[ \lambda^2 \left( D_{1} \vec{\phi} \times D_{2} \vec{\phi} \right)^2 +2 \mu^2 V(\phi \cdot \vec n) + \frac{1}{g^2} B^2 \right] \\
&=& \frac{1}{2} E_0 \int d^2 x \left[ \lambda^2 Q^2 +2 \mu^2 V(\phi_3) + \frac{1}{g^2} B^2 \right]
\end{eqnarray}
where
\begin{equation}
Q\equiv \vec \phi \cdot D_1 \vec \phi \times D_2 \vec \phi .
\end{equation}
It further holds that
\begin{equation} \label{Qq}
Q = q + \epsilon_{ij} A_i \partial_j (\vec n \cdot \vec \phi) .
\end{equation}
From now on we choose $\vec{n}=(0,0,1)$, $\vec n\cdot \vec \phi =\phi_3$. 

Next we consider the following non-negative expression
\begin{equation} \label{non-neg}
\frac{1}{2}E_0 \int d^2 x \left[ \lambda^2 (Q- w(\phi_3))^2 + \frac{1}{g^2} (B+b(\phi_3))^2 \right]
\end{equation}
where $w(\phi_3)$ and $b(\phi_3)$ are functions of $\phi_3$ which we shall determine below.  The non-negative expression may be written like
\begin{equation}
\frac{1}{2}E_0 \int d^2 x \left[ \lambda^2 Q^2+ \lambda^2 w^2 + \frac{1}{g^2} b^2 + \frac{1}{g^2} B^2 - 2\lambda^2 wq - 2\lambda^2  w \epsilon_{ij} A_i \partial_j \phi_3 + 2 \frac{1}{g^2} \epsilon_{ij} (\partial_i A_j) b \right]
\end{equation}
where we used $B=\epsilon_{ij}\partial_iA_j$.
The last two terms combine into a total derivative if we choose
\begin{equation} \label{def-b}
b(\phi_3) = g^2 \lambda^2  W(\phi_3) \equiv g^2 \lambda^2  \int_{\phi_{3,vac}}^{\phi_3} dt w(t).
\end{equation}
Indeed, the last two terms now give
\begin{equation}
E_0 \int d^2 x \lambda^2  \epsilon_{ij}\partial_j (A_i W)
\end{equation}
and this gives zero because by construction $W(\phi_3 = \phi_{3,vac}=1)=0$.
The non-negative expression therefore reads
\begin{equation}
\frac{1}{2}E_0 \int d^2 x \left[ \lambda^2 Q^2+ \lambda^2 W'^2 + g^2 \lambda^4  W^2 + \frac{1}{g^2} B^2 - 2\lambda^2 W'q  \right] .
\end{equation}
If we now require that $W$ obeys
\begin{equation} \label{AV-eq}
\lambda^2 W'^2 + g^2  \lambda^4 W^2 =2\mu^2 V
\end{equation}
then we get the BPS bound for the energy
\begin{equation} \label{BPS-bound}
E \ge E_0 \lambda^2 \int d^2 x qW' = 4\pi |k| E_0 \lambda^2 \langle W' \rangle_{S^2}
\end{equation}
where $k= {\rm deg}[\vec \phi]$ is the winding number. There is a sign ambiguity in the choice of $W$, so we may always choose the sign such that $\langle W' \rangle_{S^2} >0$. 

We remark that eq. (\ref{AV-eq}) for $W(\phi_3)$ is analogous to the "superpotential equation" which determines a superpotential $W(\phi)$ in terms of a potential $V(\phi)$ for a real scalar field  $\phi$ in the context of selfgravitating domain walls \cite{sken1} - \cite{trig1} or scalar field inflation models \cite{sken2} - \cite{bazeia3}. There the superpotential allows to reduce the domain wall or cosmological equations to a first order form. Further, the relation between potential and superpotential is completely equivalent to the relation in the corresponding (dimensionally reduced) supergravity theories, and the method is therefore called "fake supergravity". We shall find that in our case the auxiliary function $W$ allows a reduction to a first order system, as well. The main difference is that in fake supergravity the two terms proportional to $W^2$ and $W'^2$ enter with different signs in the superpotential equation, whereas they enter with the same sign in our eq. (\ref{AV-eq}), which implies some qualitative differences in the solution space, as we shall see in a moment (we remark that in the context of extremal, supersymmetric black holes the superpotential equation shows up, as well, and there both terms enter with the same sign, as in our case, see e.g. \cite{trig1}). In view of these similarities, we shall call $W$ the "superpotential" and eq. (\ref{AV-eq}) the "superpotential equation" in what follows.  

The formal energy bound (\ref{BPS-bound}) only gives a genuine BPS bound provided that either $W'$ itself or at least $\langle W' \rangle_{S^2} $ are uniquely determined. On the other hand, eq. (\ref{AV-eq}) is a first order ODE, so it seems that it provides a one-parameter family of solutions $W$. It turns out, however, that this is not true, and eq. (\ref{AV-eq}) only has one unique solution. The reason is as follows. If we want to find a local solution of eq. (\ref{AV-eq}) which is valid in the vicinity of a point $\phi_3 =c$, where $-1<c<1$, then, indeed, there exists a one-parameter family of solutions. We may choose the "initial value" $W(c)$ from the interval 
\begin{equation} \label{interval}
-\frac{\mu}{g\lambda^2} \sqrt{2V(c)} \le W(c) \le \frac{\mu}{g\lambda^2} \sqrt{2V(c)},
\end{equation}
and each choice produces one solution. The important point is that we require a solution which exists in the whole interval $-1 \le \phi_3 \le 1$, specifically at the vacuum value $\phi_3 =1$ where the potential is zero, $V(1)=0$. But for this vacuum value, the "interval" of eq. (\ref{interval})
collapses to the point $W(1)=0$, and this is the only allowed "initial value" at $\phi_3 =1$ for the differential equation (\ref{AV-eq}). The solution is, therefore, uniquely determined in the vicinity of the vacuum $\phi_3 =1$. A different (and very important) question is whether this unique solution can be extended to the whole interval $\phi_3 \in [-1,1]$, i.e., to the whole target space, see Section IV.C. Assuming this for the moment, we find for
$  \langle W' \rangle_{S^2} $ 
\begin{eqnarray}
4\pi \langle W' \rangle_{S^2} &=& \int_{S^2} d\Omega W' = \int_0^{2\pi}d\varphi \int_0^\pi d\theta \sin\theta W'(\cos \theta) \nonumber \\
&=& 2\pi \int_{-1}^1 dt W'(t) = 2\pi (W(1) - W(-1)) = -2\pi W(-1) 
\end{eqnarray}
and for the BPS bound
\begin{equation} \label{BPS-bound-2}
E \ge E_0 \lambda^2 \int d^2 x qW' = 2\pi |k| E_0  \lambda^2 |W(-1)|.
\end{equation}

\subsection{The BPS equations}
It still remains to prove that the two BPS equations imply the static field equations. The BPS equations are 
\begin{eqnarray} \label{BPS-1}
Q&=&W' \\ \label{BPS-2}
B&=&-g^2 \lambda^2  W ,
\end{eqnarray}
where we used the condition that the BPS equations hold if and only if the non-negative expression (\ref{non-neg}) is zero, together with the expression (\ref{def-b}) for $b(\phi_3)$ and $w\equiv W'$. 
On the other hand, the static second order field equations are
\begin{eqnarray} \label{s-sk-eq}
\lambda^2 \epsilon_{ij}D_i [(D_j \vec \phi ) Q] &=& -\mu^2 V' \vec n\times \vec \phi \\ \label{s-in-Max}
\partial_i F^{ij} &=& g^2 \lambda^2 \vec n \cdot D^k \vec \phi ( \vec \phi \cdot D^j \vec \phi \times D_k \vec \phi ) .
\end{eqnarray}
First we prove that the two BPS equations (\ref{BPS-1}) and (\ref{BPS-2}) imply the inhomogeneous static Maxwell equation (\ref{s-in-Max}).
With $\vec n \cdot D_k \vec \phi = \partial_k \phi_3$ the Maxwell equation reads more explicitly
\begin{equation}
\partial_k B = -g^2 \lambda^2 \partial_k \phi_3 Q.
\end{equation}
On the other hand, the partial derivative of the second BPS equation is
\begin{equation}
\partial_k B = -g^2 \lambda^2 W' \partial_k \phi_3 = -g^2 \lambda^2 Q \partial_k \phi_3
\end{equation}
where we used the first BPS equation in the last step.
It is, therefore, identical to the Maxwell equation. 

Finally we prove that the two BPS equations imply the static field equation (\ref{s-sk-eq}) for the Skyrme field. First we observe that Eq. 
(\ref{s-sk-eq}) may be re-expressed like
\begin{equation}
D_2\vec \phi \partial_1 Q - D_1 \vec \phi \partial_2 Q +  \vec n \times \vec \phi BQ = -  \lambda^{-2} \mu^2 V' \vec n\times \vec \phi .
\end{equation}
Then we use the equations
\begin{equation}
\mu^2 V' = \lambda^2 W'W'' + g^2 \lambda^4 WW' , \quad \partial_k Q = W'' \partial_k (\vec n \cdot \vec \phi) 
\end{equation}
and the second BPS equation to arrive at
\begin{equation}
(D_2 \vec \phi \partial_1 (\vec n \cdot \vec \phi ) - D_1 \vec \phi \partial_2 (\vec n \cdot \vec \phi ))W'' - g^2 \lambda^2 WW' \vec n \times \vec \phi = -(W'W'' + g^2 \lambda^2 WW')\vec n \times \vec \phi
\end{equation}
or at
\begin{equation}
D_2 \vec \phi \partial_1 (\vec n \cdot \vec \phi ) - D_1 \vec \phi \partial_2 (\vec n \cdot \vec \phi )  = -W' \vec n \times \vec \phi .
\end{equation}
Inserting for the covariant derivatives, this reads
\begin{equation}
\partial_2 \vec \phi \partial_1 (\vec n \cdot \vec \phi ) - \partial_1 \vec \phi \partial_2 (\vec n\cdot \vec \phi ) = \vec n \times \vec \phi [ A_1 \partial_2 (\vec n \cdot \vec \phi ) - A_2 \partial_1 (\vec n \cdot \vec \phi ) -W']
\end{equation}
and using a last time the BPS equation $W' = Q $ and the expression (\ref{Qq}) for $Q$  we find that all terms containing the gauge field $A_k$ disappear and we are left with
\begin{equation}
\partial_2 \vec \phi \partial_1 (\vec n \cdot \vec \phi ) - \partial_1 \vec \phi \partial_2 (\vec n\cdot \vec \phi ) = -\vec n \times \vec \phi  
(\vec \phi \cdot \partial_1 \vec \phi \times \partial_2 \vec \phi ).
\end{equation}
We still have to demonstrate tha this equation is, in fact, an identity. First, we observe that both sides in this equation are perpendicular to $\vec \phi$ and to $\vec n$, therefore we may project the only nontrivial equation by the scalar product with the vector $\vec n \times \vec \phi$ which results in
\begin{equation}
(\vec n \cdot \vec \phi \times \partial_2 \vec \phi \partial_1 (\vec n \cdot \vec \phi ) -
(\vec n \cdot \vec \phi \times \partial_1 \vec \phi \partial_2 (\vec n \cdot \vec \phi ) =
-(1-(\vec n \cdot \vec \phi )^2)\vec \phi \cdot \partial_1 \vec \phi \times \partial_2 \vec \phi .
\end{equation} 
Finally, we need the identity
\begin{equation}
\vec n \cdot \partial_1 \vec \phi \times \partial_2 \vec \phi =  (\vec n\cdot \vec \phi) \vec \phi \cdot \partial_1 \vec \phi \times \partial_2 \vec \phi
\end{equation}
which follows from the fact that $\partial_1 \vec \phi$ and  $\partial_2 \vec \phi$ span a plane perpendicular to $\vec \phi$ to arrive at
\begin{equation}
(\vec n \cdot \vec \phi \times \partial_2 \vec \phi \partial_1 (\vec n \cdot \vec \phi ) -
(\vec n \cdot \vec \phi \times \partial_1 \vec \phi \partial_2 (\vec n \cdot \vec \phi ) + 
\vec \phi \cdot \partial_1 \vec \phi \times \partial_2 \vec \phi - (\vec n \cdot \vec \phi ) 
\vec n \cdot \partial_1 \vec \phi \times \partial_2 \vec \phi =0.
\end{equation}
But this last equation is an identity as an immediate consequence of the Schouten identity in three dimensions,
\begin{equation}
\delta^{ab} \epsilon^{cde} - \delta^{ac} \epsilon^{deb} + \delta^{ad} \epsilon^{ebc} - \delta^{ae} \epsilon^{bcd} =0.
\end{equation}
Indeed, contracting the Schouten identity with $n^a n^b \phi^c \partial_1 \phi^d \partial_2 \phi^e$ we arrive at the above equation. 
Therefore, both static field equations are consequences of the BPS equations, which is what we wanted to prove.
\subsection{BPS soliton solutions}
\subsubsection{General remarks}
Eq. (\ref{AV-eq}) has one unique solution defined by the initial condition $W(1)=0, W'(1)=0$ at the vacuum value $\phi_3 =1$. If this initial condition can be integrated such that the solution covers the whole interval $-1\le \phi_3 \le 1$, then
there exists a unique, well-defined BPS bound for this model (i.e., for the corresponding potential with vacuum at $\phi_3 =1$ and for the corresponding choice of the coupling constants $\lambda$, $\mu$ and $g$). It happens that for some potentials (or coupling constants) Eq. (\ref{AV-eq}) cannot be integrated for the whole interval $-1 \le \phi_3 \le 1$. In these cases, the BPS bound derived in this section does not apply for that model. We shall find both possibilities in the following. Concretely, both for the "old" baby Skyrme potential $V_o = 1-\phi_3 =2h$ and for the potential $V_o^2 =2V_s = 4h^2$ we will find that a solution to eq. (\ref{AV-eq}) exists in the whole interval $\phi_3 \in [-1,1]$, i.e., $h\in [0,1]$. For these potentials we find that soliton solutions exist and are, in fact, solutions to the BPS equations (\ref{BPS-1}) and (\ref{BPS-2}), that is, they saturate the BPS bound (\ref{BPS-bound-2}). A global solution to eq. (\ref{AV-eq}) seems to exist for the class of potentials $V_a = (2h)^a$ (we found them by numerical integration for different values of $a$), and we believe that BPS soliton solutions exist for all values of $a$ for which eq. (\ref{AV-eq}) has a global solution. For the so-called "new" baby Skyrme potential
\be
V_n = \frac{1}{4}(1-\phi_3^2 ) = h(1-h) ,
\ee 
on the other hand, eq. (\ref{AV-eq}) does not have a global solution. Starting at $h=0$ with the initial condition $W(h=0)=0$ (and, of course, $W_h (h=0)=0$), the numerical integration develops a singularity somewhere between $h=1/2$ and $h=1$ (the precise position of the singularity depends on the values of the coupling constants $\lambda$, $\mu$ and $g$). Further, we are not able to find numerical soliton solutions for the new baby Skyrme potential, neither BPS nor non-BPS ones. These numerical findings, and the analytical results described below (see Sections IV.C.2, IV.C.3), motivate the following conjectures.
\\ \\
{\bf Conjecture 1:} If a given restricted gauged baby Skyrme model (i.e., with a given potential and fixed values of the coupling constants $\lambda$, $\mu$ and $g$) has topological soliton solutions at all, then these solutions are BPS solutions, i.e., solutions of the two BPS equations (\ref{BPS-1}), (\ref{BPS-2}), where the superpotential $W$ is the solution of the superpotential equation (\ref{AV-eq}) with boundary condition $W(\phi_3 =1) \equiv W(h=0)=0$. 
\\
{\bf Corollary 1:} The global existence of the superpotential (i.e., the existence of a solution to Eq. (\ref{AV-eq}) in the whole interval $\phi_3 \in [-1,1]$, i.e., $h \in [0,1]$) is a necessary condition for the existence of soliton solutions.
\\
{\bf Conjecture 2:} The superpotential equation (\ref{AV-eq}) and its solution $W$ completely determine the existence of soliton solutions. More specifically, in the generic case solitons exist {\em if and only if} the superpotential equation (\ref{AV-eq}) has a global solution on the whole interval $h\in [0,1]$ (i.e., on the whole target space) {\em and} this solution obeys $W' \neq 0 $ for $-1 < \phi_3 <1$ (i.e., $W_h \neq 0$ for $0<h<1$).
\\
Remark: by "generic" we mean that there might exist exceptions, i.e, potentials where the corresponding superpotential exists globally and obeys $W_h =0$ at some points but, nevertheless, supports BPS solitons. If they exist at all, these potentials will, however, be rare in the sense that they require some fine-tuning of parameters, see the discussion in the next subsection.
\\
{\bf Conjecture 3:} Potentials which have one vacuum at $\phi_3 =1$ and are strictly monotonous in the open interval $\phi_3 \in (-1,1)$ (i.e., potentials which have one vacuum at $h=0$ and are strictly monotonous in the open interval $h \in (0,1)$) satisfy the conditions of Conjecture 2 and support, therefore, topological BPS solitons.
\\
Remark: we do not think that the conditions on potentials in Conjecture 3 are necessary. That is to say, most likely there exist potentials which do not satisfy these conditions and still support BPS solitons. 
\\ 
A rigorous proof of these conjectures is probably quite difficult and is certainly beyond the scope of the present article.
\subsubsection{The problem with $W_h =0$}
We demanded in conjecture 2 not only the global existence of the superpotential $W$ but also the absence of local extrema in the open interval $0<h<1$, so let us explain the problems related to $W_h =0$. We restrict our discussion to the spherically symmetric ansatz for the BPS equations, i.e., we assume that if soliton solutions exist, at all, then they should also exist in the spherically symmetric subsector.  
We use the function $h=(1/2)(1-\phi_3)$ instead of $\phi_3$, then the superpotential equation for $W$ is 
\begin{equation} \label{AV-eq2}
\frac{\lambda^2}{4} W_h^2 + g^2 \lambda^4 W^2 = 2\mu^2 V(h)
\end{equation}
and the two BPS equations for the spherically symmetric ansatz read
\begin{equation} \label{BPS1-spher}
2nh_y (1+a) = -\frac{1}{2}W_h
\end{equation}
\begin{equation} \label{BPS2-spher}
na_y = -g^2 \lambda^2 W
\end{equation}
where $a(y)$ is the angular part of the gauge field and $y=r^2/2$. Next
we resolve  eq. (\ref{AV-eq2}) for $W_h$,
\begin{equation} \label{A_h-eq}
W_h =  \sqrt{8\frac{\mu^2}{\lambda^2} V - 4g^2 \lambda^2 W^2}
\end{equation}
and calculate the derivative
\begin{equation}
W_{hh} = \frac{4}{W_h} \left( \frac{\mu^2}{\lambda^2}V_h - g^2 \lambda^2 WW_h\right) .
\end{equation}
For $W_h=0$ this is nonsingular only provided that $V_h =0$. 
If $W_{hh}$ is singular at a point $h=h_s$ where $0<h_s <1$, then the integration breaks down at this point and cannot be extended further. The superpotential, therefore, does not exist globally, and the corresponding theory does not support BPS solitons. Further, this is the generic case in the sense that if the integration of the superpotential equation produces $W_h =0$ for some value $h_s$, then even for potentials which obey $V_h =0$ for some values $h_i$, generically $h_s$ will not coincide with any of the $h_i$, and some fine-tuning of the parameters is required to make them coincide. Specifically, if $V_h \not=0$ in the whole open interval, then $W_h =0$ automatically produces a singularity. We remark, however, that numerically we found that for such potentials $W_h =0$ never occurs, which motivated our Conjecture 3.

If $W_h =0$ occurs exactly at a point $h_s$ where $V_h =0$, too, then the superpotential may exist globally (we shall see an example in Section VI) but, still, this does not imply that BPS solitons exist. If $W_h$, e.g. has exactly one zero coinciding with one zero of $V_h$, then it follows from the first BPS equation (\ref{BPS1-spher}) that $h_y$ changes sign at the point where $W_h =0$ and is, therefore, positive either near $h=0$ or near $h=1$. But this is incompatible with the restriction $h \in [0,1]$ together with the boundary conditions $h(0)=1$, $h(\infty)=0$. It might appear that we could avoid this conclusion by assuming that it is the function $1+a$ which changes sign at $W_h =0$, but we shall see that this is impossible because $a$ satisfies the inequality $a(y)> -1 \; \forall \; y$, see Subsection IV.C.6. It follows that all $W$ where $W_h$ has an odd number of zeros in the open interval $0<h<1$ are forbidden. 

This still leaves the possibility of globally existing superpotentials with an even number of zeros of $W_h$ which support soliton solutions. At the moment we cannot exclude this possibility, but if it exists then it requires a large amount of fine-tuning. 
The derivative of the potential, $V_h$, must have at least the same number of 
zeros, and the parameters of the potential must be fine-tuned such that the positions of all the zeros of $W_h$ coincide with the positions of (some or all of) the zeros of $V_h$. This is the fine-tuning mentioned in the remark after Conjecture 2.
\subsubsection{The boundary value problem of BPS solitons}
Each solution of the BPS equations (and of the superpotential equation (\ref{AV-eq2})) is a solution of the static field equations. The converse, however, is not true in general. This is especially easy to see for the radially symmetric ansatz of (\ref{rad-ans}). There, the reason is that each BPS equation provides one integration constant and eq. (\ref{AV-eq2}) provides none, therefore there is a total of two integration constants, and the space of solutions is two-dimensional. The original static field equations, on the other hand, provide four integration constants for the radially symmetric ansatz, so their solution space is four-dimensional. \\
The important question is, of course, whether a soliton solution can be a solution of the BPS equations. A simple count of the boundary conditions which a topological soliton has to obey in the radially symmetric case seems to indicate that this is impossible, because a soliton solution has to fulfill the four boundary conditions (\ref{sol-bound-zero}), (\ref{sol-bound-comp}) in the case of compactons (or (\ref{sol-bound-zero}), (\ref{sol-bound-inf}) for non-compact solitons), which requires four integration constants. \\
What may still happen is that the condition $W(h=0)=0$, which uniquely fixes the solution of eq. (\ref{AV-eq2}), implies, at the same time, that the two boundary conditions at the compacton boundary (for compact solitons) or in the limit $y\to \infty$ (for non-compact solitons) are automatically satisfied, so that only two more boundary conditions at the center $y=0$ are left. In this case, the true soliton solution could be a BPS solution in the general case. Now we shall see that this is exactly what happens. We will explicitly discuss the case of a compact soliton, but the non-compact case is completely equivalent. Further, we shall restrict our discussion to strictly monotonous potentials where $V(h=0)$, $V_h >0$ for $0<h<1$ which requires $W_h \not=0$ in the same interval.

 It obviously follows from the first BPS equation (\ref{BPS1-spher}) that $h_y =0 \Rightarrow W_h=0$. Further, we know that the unique solution of eq. (\ref{AV-eq2}) obeys $W(h=0) = W_h (h=0) =0$ at the vacuum value $h=0$. If $h=0$ is the only point where $W_h =0$ is possible, then we can conclude that $h_y =0 \Rightarrow h=0$, which is exactly the first compacton boundary condition. But we know already that $W_h =0 $ is forbidden in the open interval $0<h<1$, therefore 
 $W_h =0$ implies either $h=0, W=0$ or $h=1, W=(\mu /g\lambda^2)\sqrt{2V(1)}$. 
\\
Further, we know that the condition $W(0)=0$, $W_h (0)=0$ at $h=0$ always holds for the solution $W$ of the superpotential equation (\ref{AV-eq2}), because it is our boundary condition for this unique solution. The condition $W_h (1)=0$, $W(1)=(\mu /g\lambda^2)\sqrt{2V(1)}$ at $h=1$, on the other hand, constitutes an additional boundary condition for this unique solution and will, therefore, hold only for exceptional, finetuned potentials. 
Specifically, it does not hold for the concrete potentials which we consider in this paper as may be checked easily by a numerical integration, therefore for these potentials we
may conclude that $h_y =0 \Rightarrow h=0$ which is exactly the first compacton boundary condition. 
\\
Equation (\ref{AV-eq2}) together with the two BPS equations (\ref{BPS1-spher}) and (\ref{BPS2-spher}), therefore, automatically imply the first compacton boundary condition ($\exists y_0 $ such that $ h_y (y_0) = h(y_0)=0$) in these cases. Further, $h(y_0)=0 \Rightarrow W(h(y_0))=0$ and the second BPS equation (\ref{BPS2-spher}) imply the second compacton boundary condition $a_y (y_0)=0$. Therefore, the two integration constants of the two BPS equations will, in general, be sufficient to fulfill the two remaining boundary conditions $h(0)=1$ and $a(0)=0$. The compacton solutions are, therefore, BPS solutions, at least for the specific class of potentials considered in this subsection. Our numerical calculations completely confirm this result.
\subsubsection{Small $g$ expansion}
\begin{figure}[h]
 \begin{center}
  \includegraphics[width=0.65\textwidth]{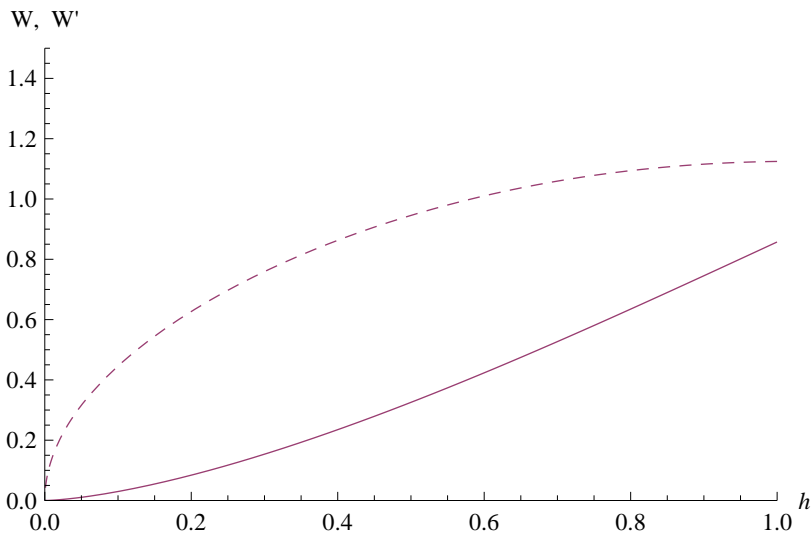}
  \caption{Solution of the superpotential equation (\ref{AV-eq2}) for the old baby Skyrme potential $V_o = 2h$, for the coupling constants $\lambda =1$, $g=1$ and $\mu^2=1/2$ ($W' \ldots $ dashed line). The solution exists in the whole interval $h\in [0,1]$.}
  \label{superpot1}
 \end{center}
\end{figure}

\begin{figure}[h]
 \begin{center}
  \includegraphics[width=0.65\textwidth]{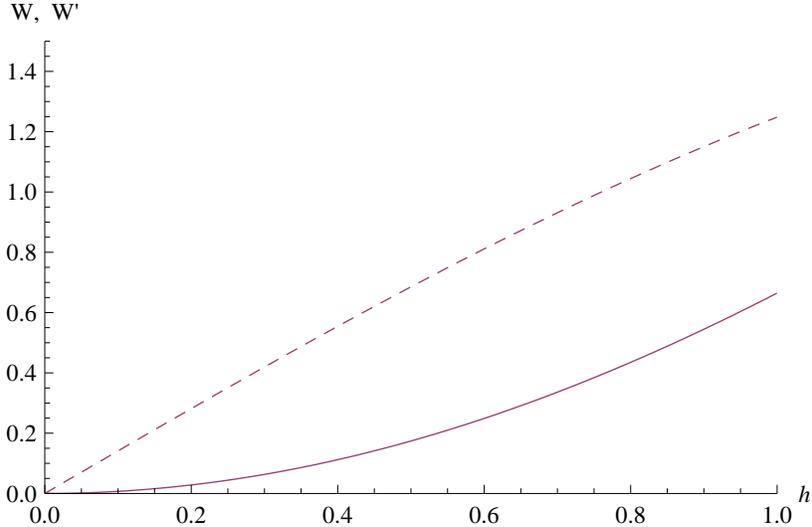}
  \caption{Solution of the superpotential equation (\ref{AV-eq2}) for the potential $V_s = 2h^2$, for the coupling constants $\lambda =1$, $g=1$ and $\mu^2=1/2$ ($W' \ldots $ dashed line). The solution exists in the whole interval $h\in [0,1]$.}
  \label{superpot2}
 \end{center}
\end{figure}

We mentioned already that the specific potentials $V_a \sim h^a $ all seem to allow for a global solution to the superpotential equation and, therefore, for a BPS bound. We show the result of a numerical integration for the values $a=1$, i.e., the old baby Skyrme potential, and for $a=2$, i.e., for the potential $V_s = 2h^2$, in Figures \ref{superpot1} and \ref{superpot2}, respectively. If a BPS bound exists, then in the limit of vanishing  electromagnetic coupling $g$ the resulting BPS bound is just the BPS bound of the ungauged BPS baby Skyrme model, and there exists an exact expansion for Eq. (\ref{AV-eq2}) in $g^2$. We may, therefore, use this power series expansion to get exact (instead of just numerical) values for the BPS bound.  Concretely, the power series expansion for $W$ may be found from eq. (\ref{A_h-eq}), re-expressed as
\begin{equation}
W_h = \frac{2\mu}{\lambda}\sqrt{2V} \sqrt{ 1- \frac{g^2 \lambda^4}{2\mu^2}\frac{W^2}{V}}
\end{equation}
by the following three-step process. i) solve the equation iteratively for $W_h$, ii) expand the second root in a power series in $g^2$ and iii) integrate the resulting power series w.r.t. $h$, respecting the boundary condition $W(h=0)=0$. Let us consider the class of potentials $V=h^{2\alpha}$, $\alpha >0$, as an example. Applying the procedure, we easily find that up to first order in $g^2$ the resulting $W$ reads
\begin{equation}
W^{(1)} (h) = \frac{2\sqrt{2}\mu}{\lambda (\alpha + 1)} h^{\alpha + 1} \left( 1 - \frac{\lambda^2}{(\alpha + 1)(\alpha + 3)} g^2 h^2 \right) .
\end{equation}
Evaluating this expression at $h=1$ gives the BPS bound. Here, the leading order $g^0$ is the bound of the ungauged model, and the leading correction for small $g$ is of order $g^2$ and negative. All these results, including the precise numerical values, are confirmed by our numerical calculations. On the other hand, there does not seem to exist an expansion of the superpotential equation for large $g$. Numerically, we find that the BPS bound (and, therefore, also the soliton energy) for large $g$ behaves like $g^{-1}$ in the case of the old baby Skyrme potential, see Section V. 
\subsubsection{Potentials with two vacua}
\begin{figure}[h]
 \begin{center}
  \includegraphics[width=0.65\textwidth]{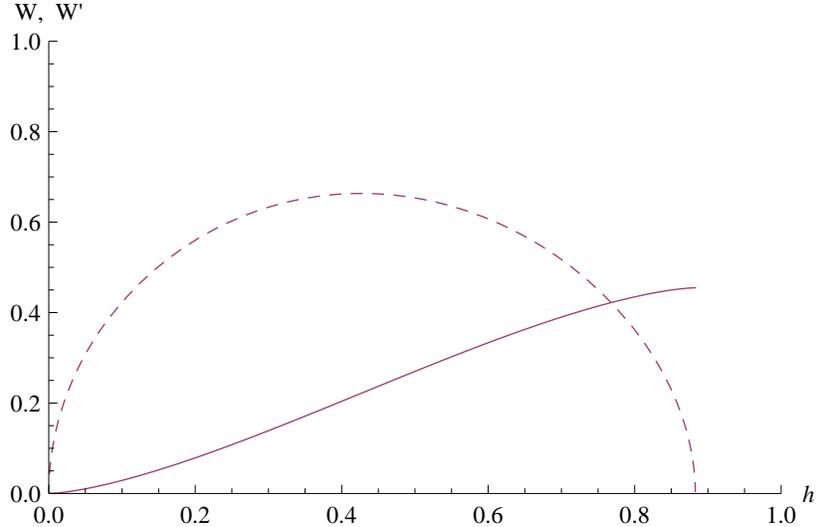}
  \caption{Solution of the superpotential equation (\ref{AV-eq2}) for the new baby Skyrme potential $V_n = h(1-h)$, for the coupling constants $\lambda =1$, $g=1$ and $\mu =1$ ($W' \ldots $ dashed line). The solution develops a singularity at $h\sim 0.9$ and cannot be extended to the whole interval $h \in [0,1]$.}
  \label{superpot3}
 \end{center}
\end{figure}

Here we want to show that for potentials with two vacua BPS soliton solutions cannot exist. Concretely we shall focus on potentials with their two vacua at $h=0,1$, although the generalization to other cases poses no difficulty. One specific example is given by the
new baby Skyrme potential $V_n =h(1-h)$, but the same arguments apply for the general case. It is easy to understand why generically the superpotential equation does not have a global solution. The problem is that, as the potential vanishes at the two points $h=0$ and $h=1$, the putative solution would have to obey the two boundary conditions $W(0)=0$ and $W(1)=0$, but these are too many conditions for a first order equation. 
 We show the result of a numerical integration for the new baby Skyrme potential in Figure \ref{superpot3}, where the singularity is clearly visible.
As already mentioned in Subsection IV.C.1, for the new baby Skyrme potential there do not seem to exist soliton solutions, at all. It is not surprising that there are no BPS soliton solutions, because we need the superpotential in the BPS equations, so if there is no global superpotential solution we do not expect true (globally existing) BPS soliton solutions. On the other hand, it is not so obvious why there are no soliton solutions, at all (i.e., neither BPS nor non-BPS soliton solutions), although we found some indications that all soliton solutions are, in fact, BPS solutions (see the results of Subsection IV.C.3). The non-existence of solitons is in some sense surprising because the non-gauged BPS baby Skyrme model (i.e., the case $g=0$) with the new baby Skyrme potential has a perfectly well-defined BPS bound and soliton solutions saturating this bound. Nevertheless, both the BPS bound and the soliton solutions cease to exist for arbitrarily small but nonzero gauge coupling $g$. The resolution of this puzzle resides, of course, in the superpotential equation which tells us that the superpotential $W$ does not have to satisfy any boundary condition in the non-gauge case with $g$ strictly zero. On the other hand, for arbitrarily small but nonzero $g$, $W$ has to obey the two boundary conditions $W(h=0)=W(h=1)=0$, which is impossible. It would be quite difficult to understand this result (the non-existence of a BPS bound and of solitons in the gauged model) without the additional insight provided by the superpotential equation.

The case of the new baby Skyrme potential where the two boundary conditions $W(h=0)=0$ and $W(h=1)=0$ cannot be satisfied simultaneously corresponds to the generic case of two-vacua potentials. Still, there will exist some fine-tuned potentials for which the two boundary conditions can be satisfied (we shall see an explicit example in Section VI). These are precisely the fine-tuned cases where the extremum of $V$  - which must exist because $V$ has two vacua - coincides with the extremum of $W$ - which must exist because $W$ interpolates between $W(h=0)=0$ and $W(h=1)=0$. It follows from the general results of subsection IV.C.2 that also in these cases soliton solutions do not exist. We conclude that BPS solitons cannot exist for potentials with two vacua.
\subsubsection{The magnetic flux}
Here we want to demonstrate that the magnetic flux of a spherically symmetric soliton can be expressed in terms of the superpotential. Therefore, if we know the superpotential then we know the magnetic flux exactly. In the course of the derivation we shall also find the inequality $a(y) > -1 \; \forall \; y$. 
The magnetic flux may be expressed in terms of the asymptotic value of $a$ like
\be
\Phi = \int rdrd\varphi B = 2\pi n \int dy a_y = 2\pi n a(y_0) \equiv 2\pi n a_\infty
\ee
where $y_0$ is finite for compactons and infinite for non-compact solitons; further, $n$ is the winding number of the spherically symmetric ansatz. 
Dividing the second BPS equation (\ref{BPS2-spher}) by the first (\ref{BPS1-spher}) we find
\be
\frac{a_y}{1+a} = 4g^2 \lambda^2 h_y \frac{W}{W_h}
\ee
or
\be
\partial_y \ln (1+a) = g^2 \lambda^2 \partial_y F
\ee
where 
\be
F_h \equiv 4\frac{W}{W_h} \quad \Rightarrow \quad F (h) = 4\int_0^h dh' \frac{W(h')}{W_h (h')}
\ee
which leads to the $y$ integral
\be
\ln C(1+a) = g^2 \lambda^2 F(h(y)) 
\ee
where $C$ is an integration constant. Here we assumed that the potential is generic, i.e., that $W_h =0$ does  not occur in the interval $0<h\le 1$, which implies that $F_h$ is finite in the same interval. At the vacuum $h=0$, where $W_h =0$, we assume that the potential behaves algebraically, i.e. $V \sim h^{2\alpha}$ for some $\alpha >0$, then $W_h \sim h^\alpha $, $W \sim h^{\alpha +1}$ near $h=0$ and $F_h$ is, in fact, zero at $h=0$. As a consequence, $F$ exists and is finite in the whole interval $h\in [0,1]$. For these generic potentials, it follows from the above result that 
\be
a(y) > -1 \quad \forall \quad y 
\ee
and that the limit $a\to -1$ may be reached only in the limit $g\lambda \to \infty$. The integration constant may be determined from the boundary conditions $h(y=0)=1$, $a(y=0)=0$,
\be
F(1)=\frac{1}{g^2\lambda^2} \ln C \quad \Rightarrow \quad C=e^{g^2 \lambda^2 F(1)}
\ee
which, together with $h(y_0)=0$ and $F(h=0)=0$ leads to the asymptotic expression
\be
a_\infty =-1+e^{-g^2\lambda^2 F(1)}
\ee
which may be inserted into the expression for the magnetic flux. Specifically, in the limits of small and large electromagnetic coupling $g$ we find for the magnetic flux
\be
g \quad \mbox{small}: \; \Phi \sim -2\pi n g^2\lambda^2 F(1)
\ee
\be
g \quad \mbox{large}: \; \Phi \sim - 2\pi n .
\ee
Both the small and large $g$ behaviour coincide with the numerical findings for the full gauged baby Skyrme model in \cite{GPS}, but here it is an exact result. We remark that this result is completely confirmed by our numerical calculations, as well.

\section{Numerical solutions}
In this section we present the results of numerically solving the static field equations within the spherically symmetric ansatz (\ref{h-eq}), (\ref{a-eq}). Concretely, we will consider the two cases of the old baby Skyrme potential $V_o = 2h$ and of the potential $V_o^2 = 2V_s = 4h^2$. 
\subsection{The old baby Skyrme potential}
The static field equations for the old baby Skyrme potential with the spherically symmetric ansatz allow for a first integral which leads to
the following equations (see Section 2.4; $y=r^2 /2$)
\begin{equation} \label{old-stateq1}
 h_y (1+a)^2 = \frac{\mu^2}{2 n^2 \lambda^2} (y-y_0),
\end{equation}
\begin{equation} \label{old-stateq2}
 (1+a)^3 a_{yy} = \lambda^2 g^2 \left( \frac{\mu^2}{n^2 \lambda^2} \right)^2 (y-y_0)^2,
\end{equation}
where $y_0$ is the integration constant of the first integral. Physically, $y_0$ is interpreted as the (squared) compacton radius. Further, the energy density is
\begin{equation}
 \epsilon=2 \lambda^2 n^2 (1+a)^2 h_y^2 + 2 \mu^2 h + \frac{1}{2g^2} n^2 a_y^2,
\end{equation}
and, taking into account the change of variable, $dy = rdr$, the static energy is (we choose the energy scale $E_0 =1$)
\begin{equation}
 E = \int dy \epsilon(y).
\end{equation}
Obviously, the static field equations depend on the topological charge $n$ and on the coupling constants $\lambda$ and $\mu$ only via the combination $(\mu /n \lambda )$, therefore we may fix two of them and just vary the third one. Concretely we will choose $n=1$ and $\lambda =1$, where the second choice just fixes our length scale. With this choice, $\mu$ is a dimensionless parameter (coupling constant), and different values of $\mu$ correspond to different theories.

\begin{figure}[h]
 \begin{center}
  \subfloat[Function $h$ and its derivative (dashed line).]{\includegraphics[width=0.6\textwidth]{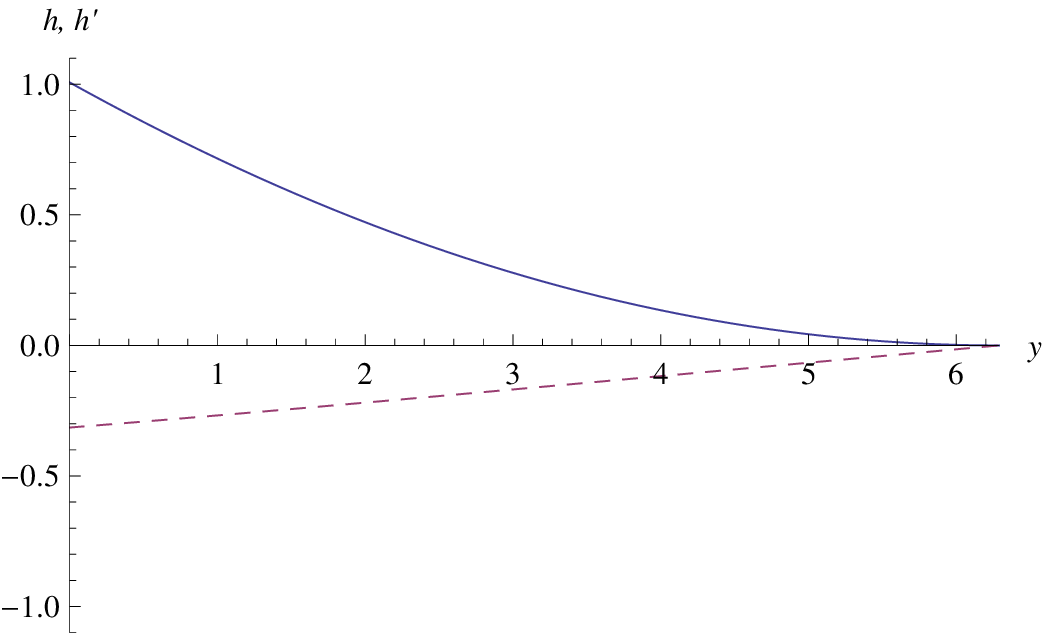}} \\
  \subfloat[Function $a$ and its derivative (dashed line).]{\includegraphics[width=0.6\textwidth]{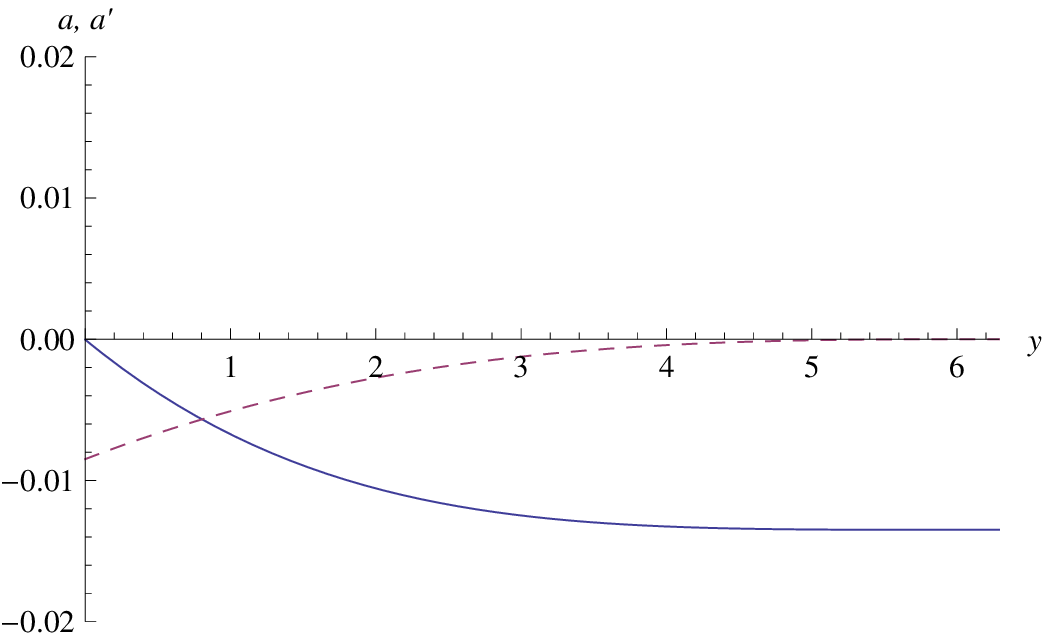}} \\
  \subfloat[Energy density.]{\includegraphics[width=0.6\textwidth]{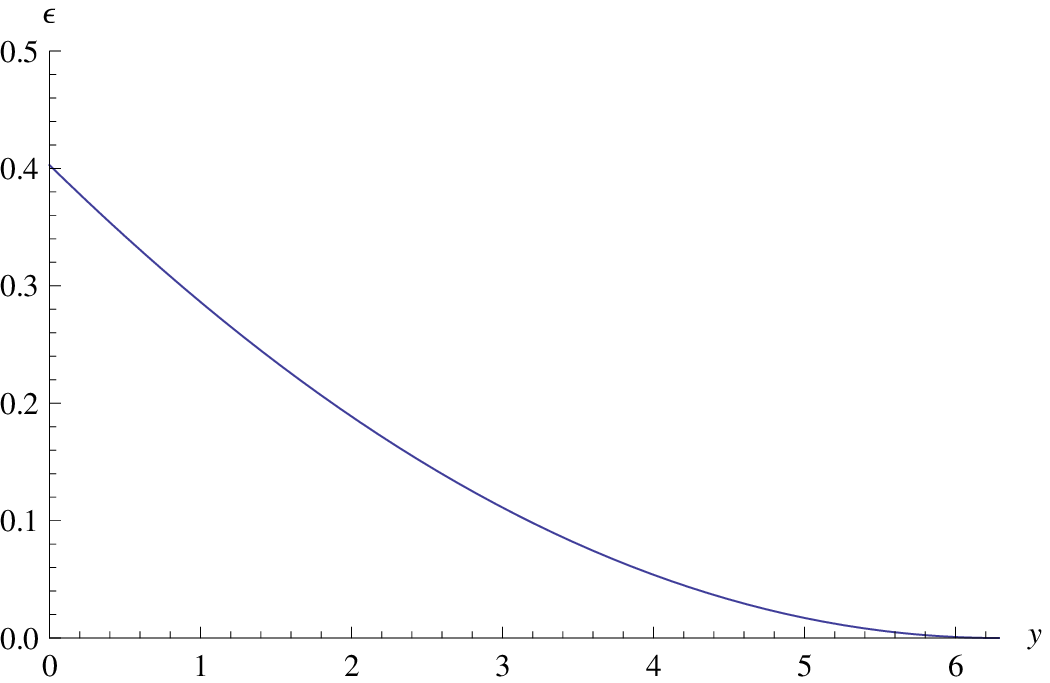}} \\
  \caption{Solutions for $g=0.1$.}
  \label{g01}
 \end{center}
\end{figure}

\begin{figure}[h]
 \begin{center}
  \subfloat[Function $h$ and its derivative (dashed line).]{\includegraphics[width=0.6\textwidth]{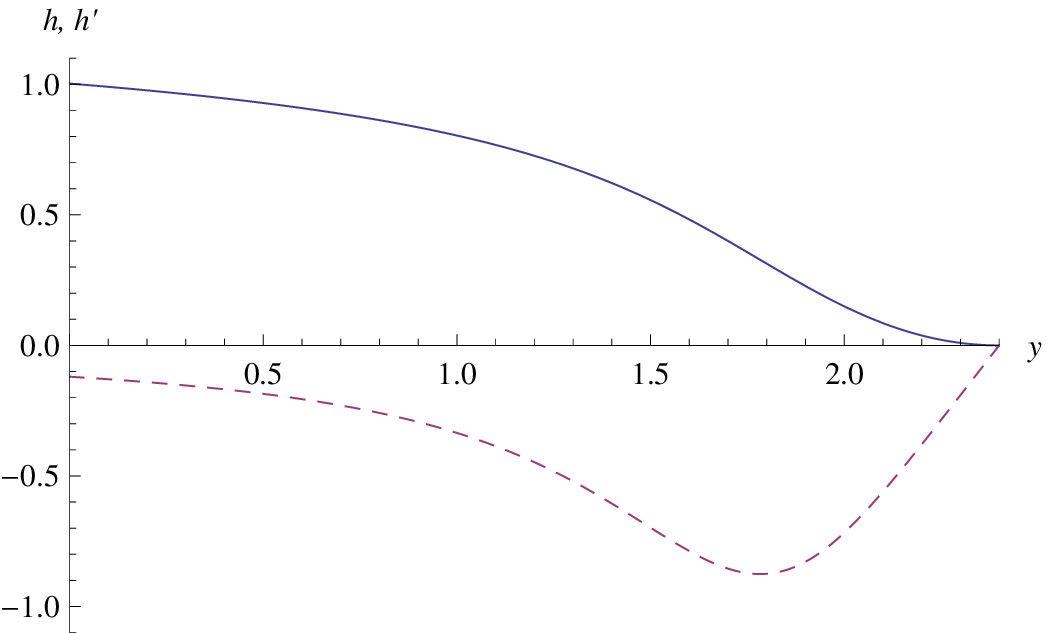}} \\
  \subfloat[Function $a$ and its derivative (dashed line).]{\includegraphics[width=0.6\textwidth]{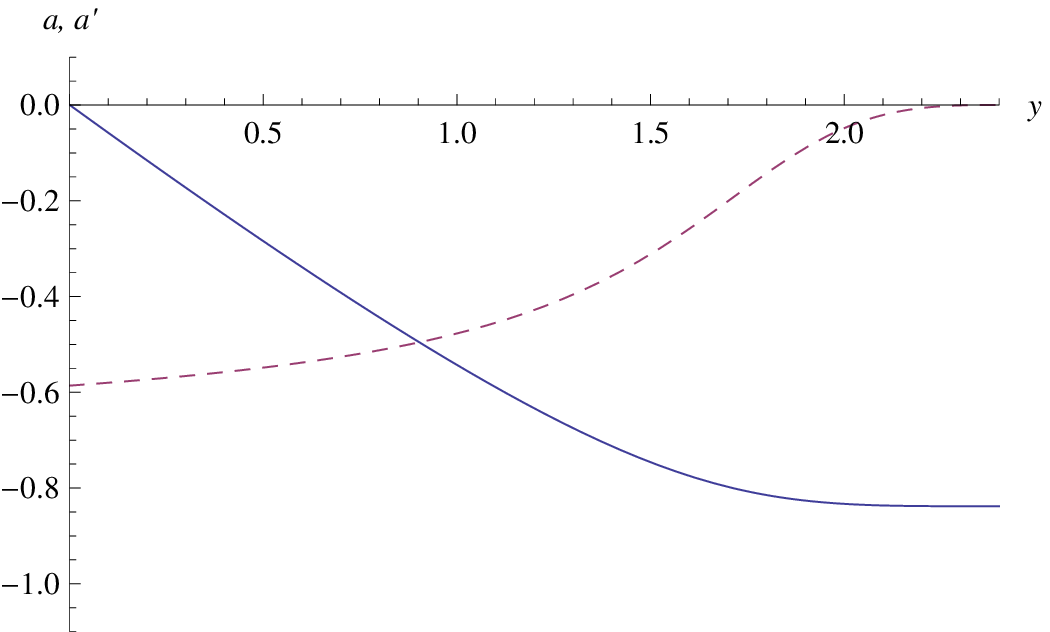}} \\
  \subfloat[Energy density.]{\includegraphics[width=0.6\textwidth]{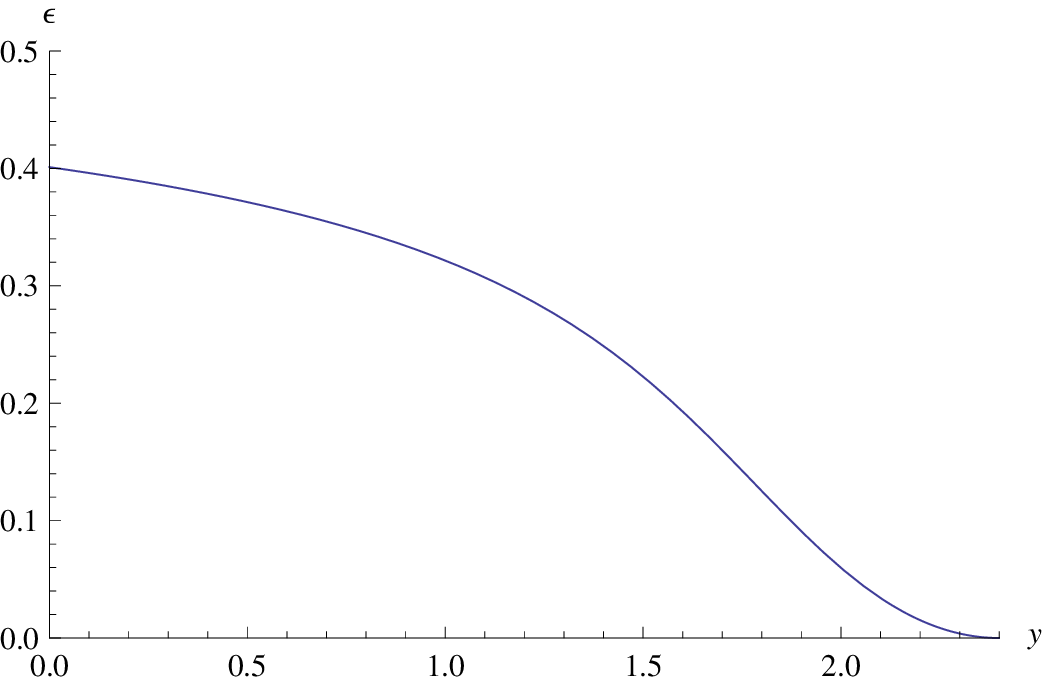}} \\
  \caption{Solutions for $g=1$.}
  \label{g1}
 \end{center}
\end{figure}

\begin{figure}[h]
 \begin{center}
  \subfloat[Function $h$ and its derivative (dashed line).]{\includegraphics[width=0.6\textwidth]{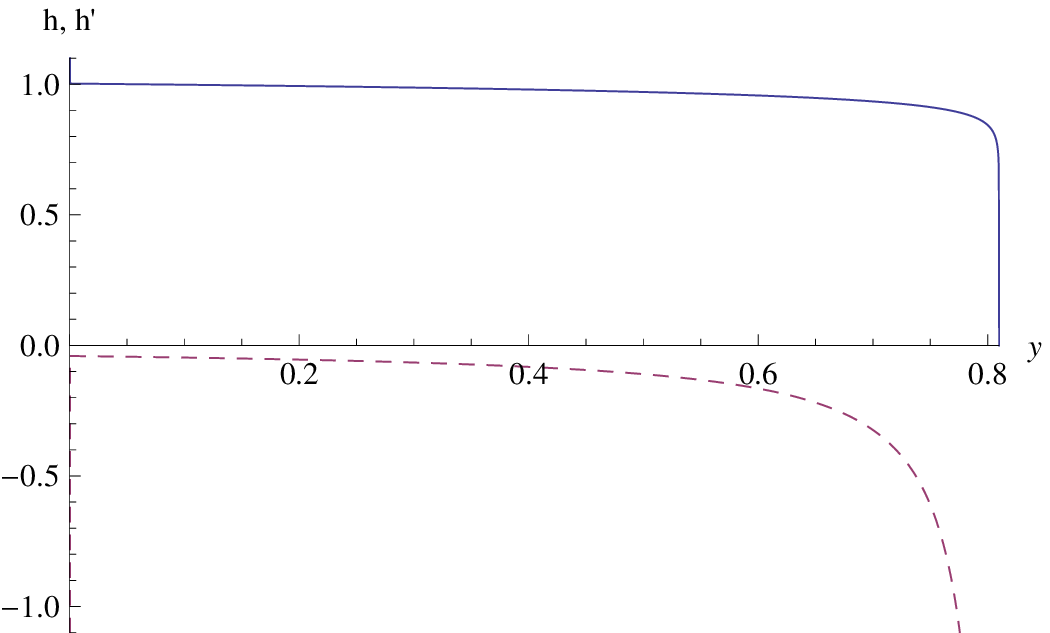}} \\
  \subfloat[Function $a$ and its derivative (dashed line).]{\includegraphics[width=0.6\textwidth]{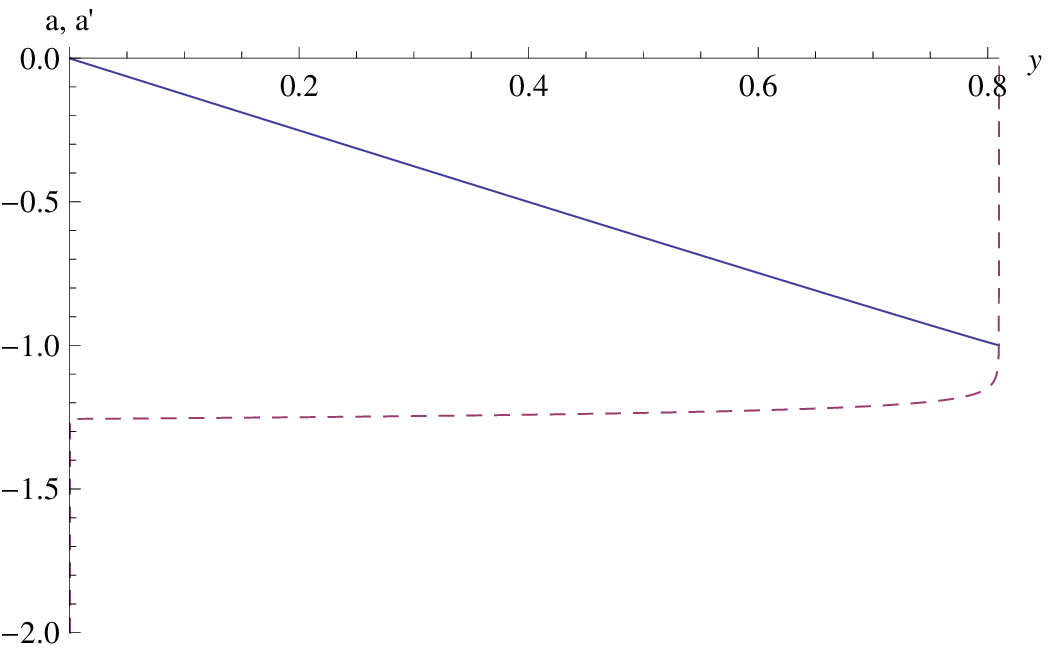}} \\
  \subfloat[Energy density.]{\includegraphics[width=0.6\textwidth]{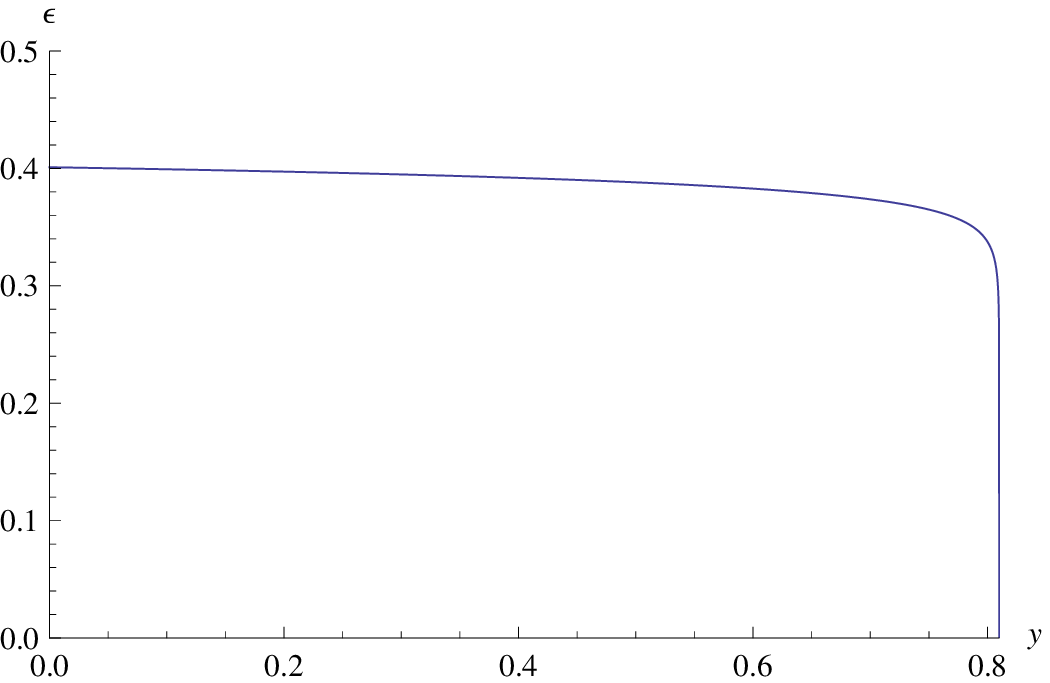}} \\
  \caption{Solutions for $g=2$.}
  \label{g2}
 \end{center}
\end{figure}

\begin{figure}[h]
 \begin{center}
  \subfloat[Function $h$ (magnified near the compacton boundary) and its derivative (dashed line).]{\includegraphics[width=0.6\textwidth]{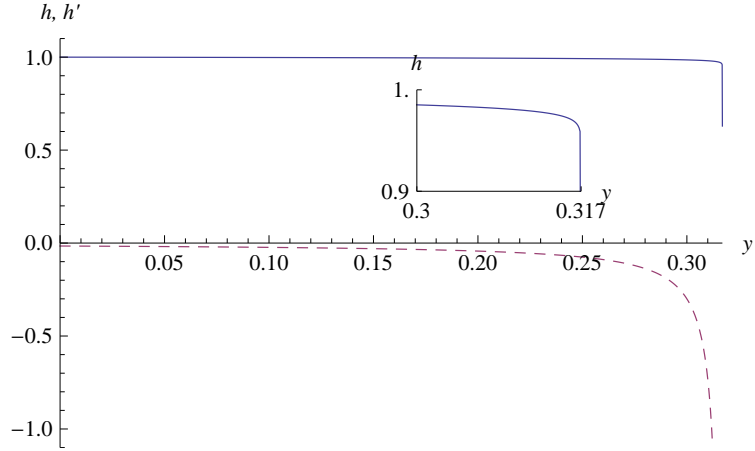}} \\
  \subfloat[Function $a$ and its derivative (rescaled by $10^{-1}$; dashed line).]{\includegraphics[width=0.6\textwidth]{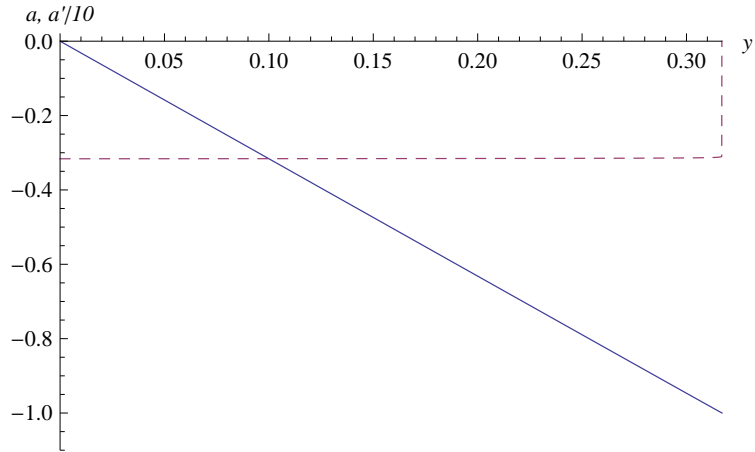}} \\
  \subfloat[Energy density (magnified near the compacton boundary).]{\includegraphics[width=0.6\textwidth]{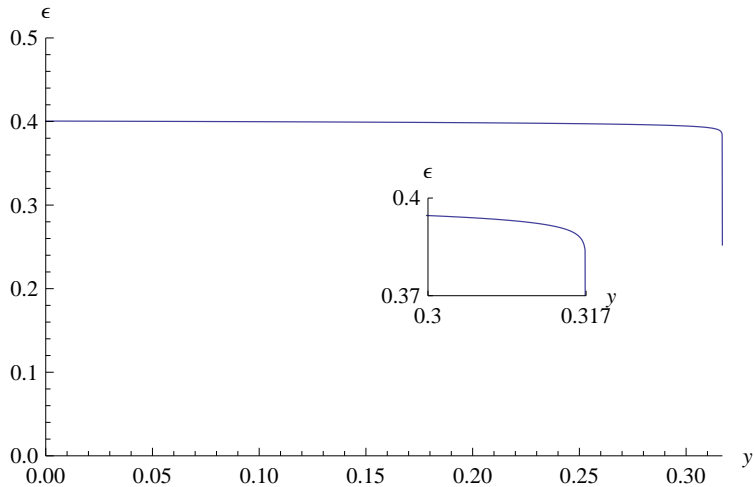}} \\
  \caption{Solutions for $g=5$.}
  \label{g5}
 \end{center}
\end{figure}

\begin{figure}[h]
 \begin{center}
  \subfloat[Function $h$ (magnified near the compacton boundary) and its derivative (dashed line).]{\includegraphics[width=0.6\textwidth]{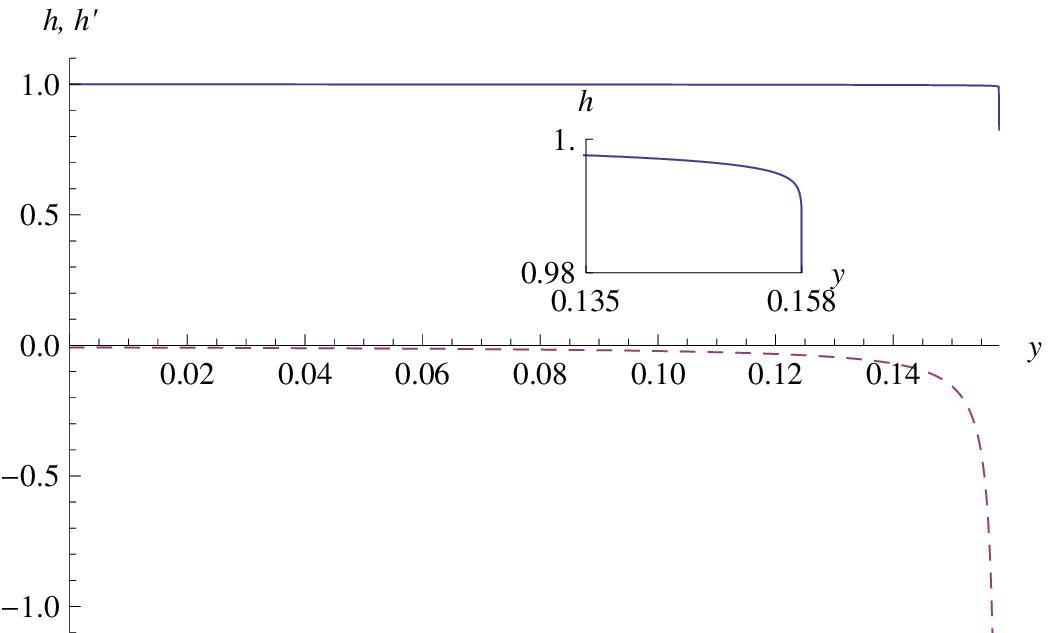}} \\
  \subfloat[Function $a$ and its derivative (rescaled by $10^{-1}$; dashed line).]{\includegraphics[width=0.6\textwidth]{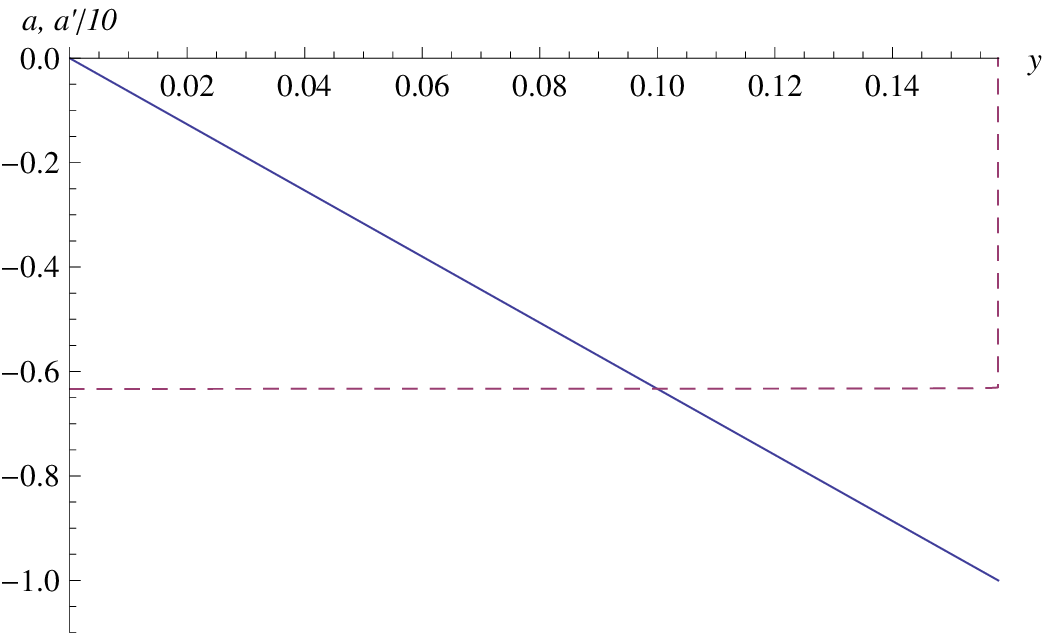}} \\
  \subfloat[Energy density (magnified near the compacton boundary).]{\includegraphics[width=0.6\textwidth]{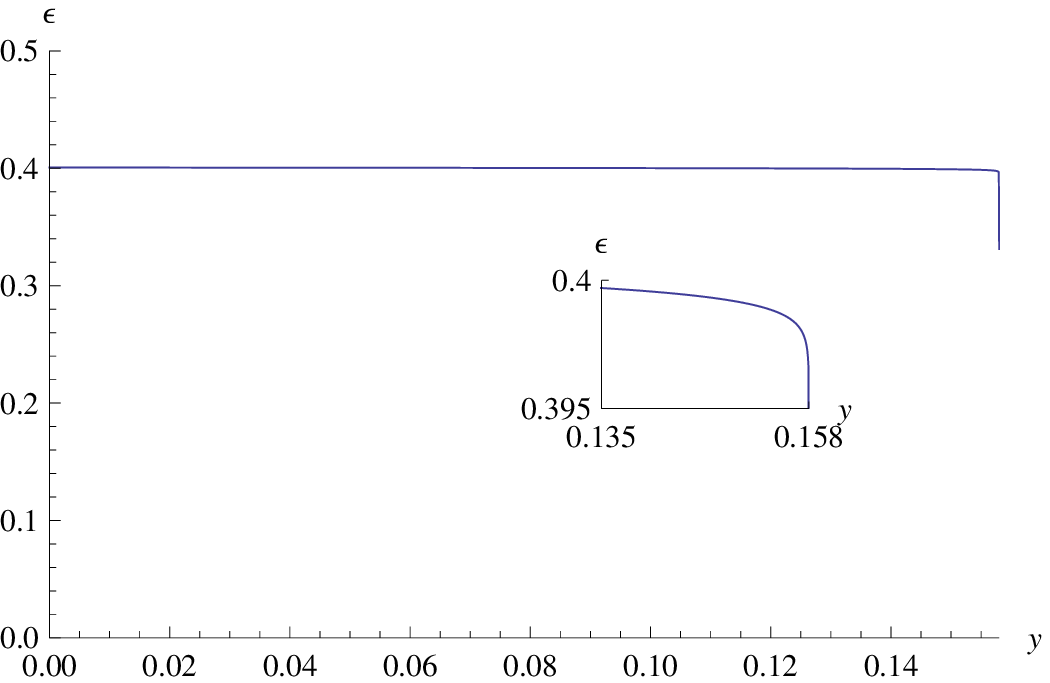}} \\
  \caption{Solutions for $g=10$.}
  \label{g10}
 \end{center}
\end{figure}

Before performing the numerical calculations, it is useful to do a power series expansion about the compacton boundary. Inserting the expansion into the static field equations we find in leading order
\begin{equation}
 h = \frac{\mu^2}{4 n^2 \lambda^2 (1+b_0)^2} (y-y_0)^2 + O(g^2 (y-y_0)^6),
\end{equation}
\begin{equation}
 a = b_0 + \frac{g^2 \mu^4}{12 n^4 \lambda^2 (1+b_0)^3} (y-y_0)^4 + O(g^4 (y-y_0)^8).
\end{equation}
Here, $b_0$ is a free parameter (i.e., it remains undetermined by the equations). If we perform a shooting from the boundary, we have therefore the two free parameters $y_0$ and $b_0$ at our disposal which we may vary in order to satisfy the two remaining boundary conditions $h(y=0)=1$ and $a(y=0)=0$. We have performed the shooting from the compacton boundary for the values $\mu^2$ ($\mu^2= 0.1, 1, 10, 50$) with very similar results, so here we will only show 
the first one, $\mu^2 = 0.1$. The main reason for this choice $\mu^2 =0.1$ is that this is the value chosen in reference \cite{GPS}. 
We have also chosen different values of the coupling constant $g$. Thus, in Table \ref{table1}  we
show the parameter values and energies for the solutions found for the values ($g = 0.001, 0.01, 0.1, 1, 2$). Further, we show the graphs of the solutions in Figures \ref{g01} - \ref{g2} 
for the values $g=0.1,1,2$. We do  not display the figures for $g=0.001, 0.01$ because they look exactly like Fig. \ref{g01} for $g=0.1$, with the only difference that the graph for the gauge field $a$ (and its derivative $a'$) has to be multiplied by $10^{-2}$ for $g=0.01$ and by $10^{-4}$ for $g=0.001$, because $a$ is proportional to $g^2$ for small $g$ to a high precision. 
\begin{center}
\begin{table}
 \begin{tabular}{|c|c|c|c|c|}
 \hline
  {\it g} & $y_0$ & $b_0$ & $E$ & Figure \\
 \hline
  0.001 & 6.325 & $-1.1361 \cdot 10^{-6}$ & 5.2996 & (\ref{g01}) \\
 \hline
  0.01 & 6.324 & $-0.134 \cdot 10^{-3}$ & 5.2983 & (\ref{g01}) \\
 \hline
  0.1 & 6.268 & -0.0135 & 5.2794 & \ref{g01} \\
 \hline
  1 & 2.397 & -0.838 & 3.6760 & \ref{g1} \\
 \hline
  2 & 0.810 & -0.9999964 & 1.9711 & \ref{g2} \\
  \hline
\end{tabular}
\caption{Solutions of field equations for $\mu^2=0.1$ and low $g$.}
\label{table1}
\end{table}
\end{center}
On the other hand, for large values of $g$ shooting from the boundary is problematic for the following reason. We cannot start the shooting exactly at the compacton boundary, because the fields take exactly their vacuum values at the boundary, and the numerical integration would only find the trivial vacuum solution and not the soliton. Instead, we have to start the integration slightly inside the compacton radius and use the above power series expansion for the determination of the "initial" value (i.e., boundary value). The problem is that this power series expansion in $y-y_0$ is, at the same time, a power series expansion in $g$. It is therefore reliable for small $g$ but not for large $g$. The way out is to perform a shooting from the center for large $g$. A power series expansion in $y$ about the center $y=0$ is again, at the same time, a power series expansion in $g$. The difference is that the fields do not take their vacuum values at the center, therefore we may start the shooting exactly at the center.  Inserting the power series expansion at the center into the field equations, we get
\begin{equation}
h(y) \sim 1 - \frac{\mu^2 y_0}{2 n^2 \lambda^2} y + ... \, ,
\end{equation}
\begin{equation}
a(y) \sim b_1 y + \frac{g^2 \mu^4 y_0^2}{2 n^2 \lambda^2} y^2 + ... \,
\end{equation}
 so we have the free parameters $b_1$ and $y_0$ to satisfy the two boundary conditions at the compacton boundary. We show the solutions for $g=5,10$ in Table \ref{table1bis} and in Figures \ref{g5} and \ref{g10}.

\begin{table}
\begin{center}
 \begin{tabular}{|c|c|c|c|c|}
 \hline
  {\it g} & $y_0$ & $b_1$ & $E$ & Figure \\
\hline
 5 & 0.317 &  -3.16261 & 0.793768 & \ref{g5} \\
\hline
10 & 0.158 & -6.33307 & 0.397260 & \ref{g10} \\
  \hline
\end{tabular}
\end{center}
\caption{Solutions of field equations for $\mu^2=0.1$ and high $g$.}
\label{table1bis}
\end{table}

\begin{figure}[h]
 \begin{center}
  \includegraphics[width=0.6\textwidth]{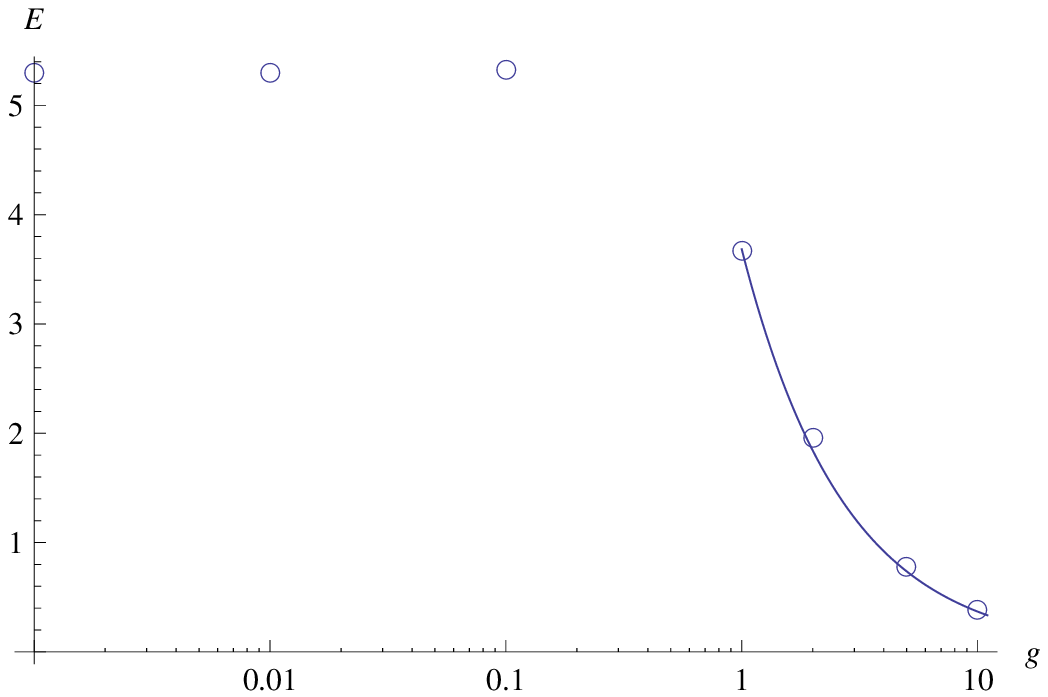}
  \caption{Static energy as a function of $g$ for $\mu^2=0.1$. The continuous line shows the behaviour $1/g$ for $g \geq 1$. We have found a similar behaviour for the other values of $\mu^2$.}
  \label{e_g}
 \end{center}
\end{figure}

\begin{figure}[h]
\begin{center}
\includegraphics[width=0.6\textwidth]{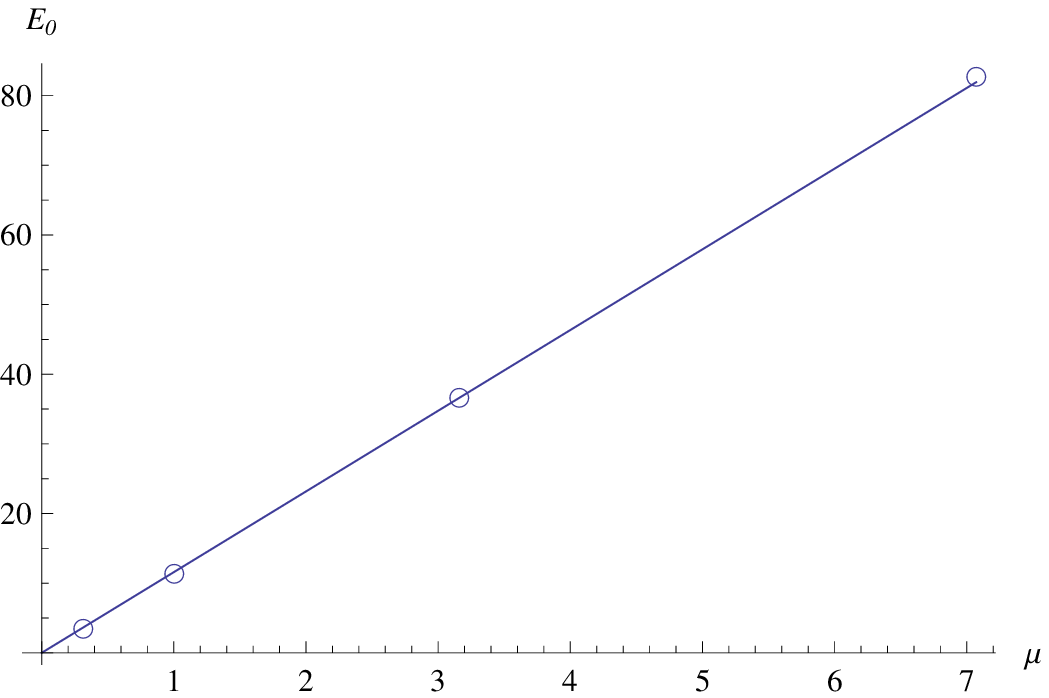}
\caption{Static energy as a function of $\mu$ for $g=0.1$. The continuous line shows the behaviour proportional to $\mu$. We have found similar plots for the other values of $g$.}
\label{e_mu}
\end{center}
\end{figure}

Now let us briefly comment on the behaviour of the soliton energies as a function of the gauge coupling constant $g$. We find numerically that 
 the energies behave as $1/g$ for $g \geq 1$ whereas they are more or less constant when $g<1$, for all values of $\mu^2$ studied. The behaviour for small $g$ exactly reproduces the analytical result found in Section IV.C, Remark 3. On the other hand, we were not able to find an analytic expression for the large $g$ behaviour to compare with. The behaviour for different values of $g$ for the constant value $\mu^2=0.1$ is presented in Figure \ref{e_g}. In addition, studying the variation of the energy with $\mu$ for a fixed value of $g$, we find that it is proportional to $\mu$ (see Figure \ref{e_mu} where $g = 0.1$).

Another quantity of considerable physical interest is the magnetic flux
\be
\Phi = \int rdrd\varphi B = 2\pi n \int dy a_y = 2\pi n a(y_0) = 2\pi n b_0
\ee
where in the case of the shooting from the boundary the magnetic flux may be expressed directly in terms of the free integration constant $b_0$. We may infer from Table \ref{table1}  that the magnetic flux grows like $g^2$ to a high precision for small $g$, whereas it approaches the constant, "quantized" value $-2\pi n$ for large $g$. That is to say, the behaviour we find for the magnetic flux reproduces exactly the analytical results of Section VI.C.6 and coincides with the one found in \cite{GPS} for the full gauged baby Skyrme model. The behaviour of the magnetic field itself, on the other hand, is different from the results of \cite{GPS}. We find that, for large $g$, $a(y)$ changes almost exactly linearly from $a(0)=0$ to $a(y_0)\simeq-1$. The magnetic field $B=na_r /r = na_y$ is, therefore, almost constant inside the compacton and rapidly decreases to zero near the compacton boundary $y=y_0$ (we remark that the (almost constant) value of $a_y$ for large $g$ is too large to fit into the figures \ref{g5}, \ref{g10}, therefore we rescaled it by $1/10$). Further, the compacton radius squared $y_0$ shrinks like $g^{-1}$ for large $g$, so the compacton radius $r_0 = \sqrt{2y_0}$ shrinks like $g^{-1/2}$, and the (constant) magnetic field inside the compacton grows like $g$. This should be contrasted with the findings of \cite{GPS}, where they find a magnetic field which is almost completely concentrated in a tiny region about $y=0$, i.e., $r=0$ for large $g$. 

\subsection{Integrating the BPS equations}
\begin{figure}[h]
\begin{center}
\includegraphics[width=0.6\textwidth]{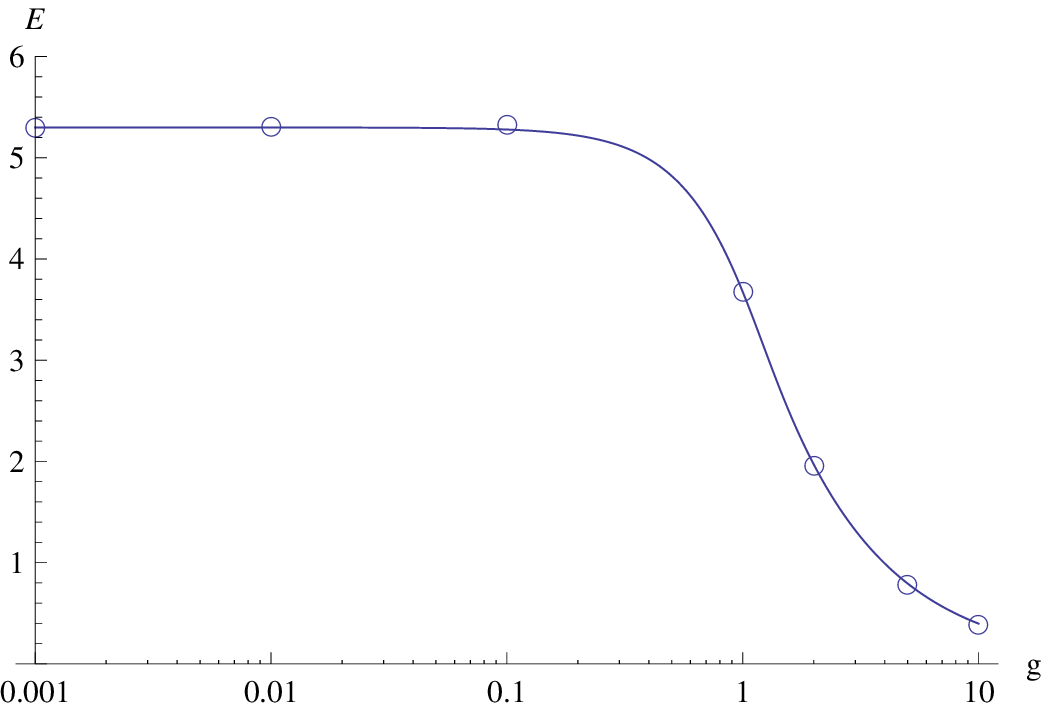}
\caption{Energy of static solutions (open circles) compared to the BPS bound (continuous line).}
\label{EB_E0}
\end{center}
\end{figure}

In a next step, we want to compare the energies and solutions of the static field equations (\ref{old-stateq1}), (\ref{old-stateq2}) with the energies and solutions of the BPS and superpotential equations, in order to confirm that the solitons are, indeed, BPS solutions. In a first step, we compare the soliton energies with the BPS bound, where for the BPS bound we just have to determine $W(h=1)$, i.e., $W(\phi_3 =-1)$ numerically for the old baby Skyrme potential, see eq. (\ref{BPS-bound-2}). 
We have done this for different values of $\mu^2$ with similar results, although, as before, here we only show the case $\mu^2=0.1$. Then, in Table \ref{table2} we show the values of the BPS energy, $E_B$, for each value of $g$, comparing them to the energies of our compacton solutions. In addition, in Figure \ref{EB_E0} both energies are presented.

\begin{table}
\begin{center}
 \begin{tabular}{|c|c|c|}
 \hline
  {\it g} & $E_B$ & $E$  \\
 \hline
  0.001 & 5.2984 & 5.2996 \\
 \hline
  0.01 & 5.2982 & 5.2983 \\
 \hline
  0.1 & 5.2783 & 5.2794 \\
 \hline
  1 & 3.6744 & 3.6760 \\
 \hline
  2 & 1.9705 & 1.9711 \\
  \hline
  5 & 0.793765 & 0.793768 \\
  \hline
  10 & 0.397259 & 0.397260 \\
  \hline
\end{tabular}
\end{center}
\caption{BPS bound and static soliton energies for $\mu^2 = 0.1$.}
\label{table2}
\end{table}
The values of the BPS energies and the energies of our solutions agree with a precision of better than $5\cdot 10^{-4}$ in all cases, and all numerical energies are slightly above the BPS energies, as obviously must be true.  We also remark that the values of $E_B$ obtained for small $g$ follow exactly the analytical expression.

\begin{figure}[h]
 \begin{center}
  \includegraphics[width=0.6\textwidth]{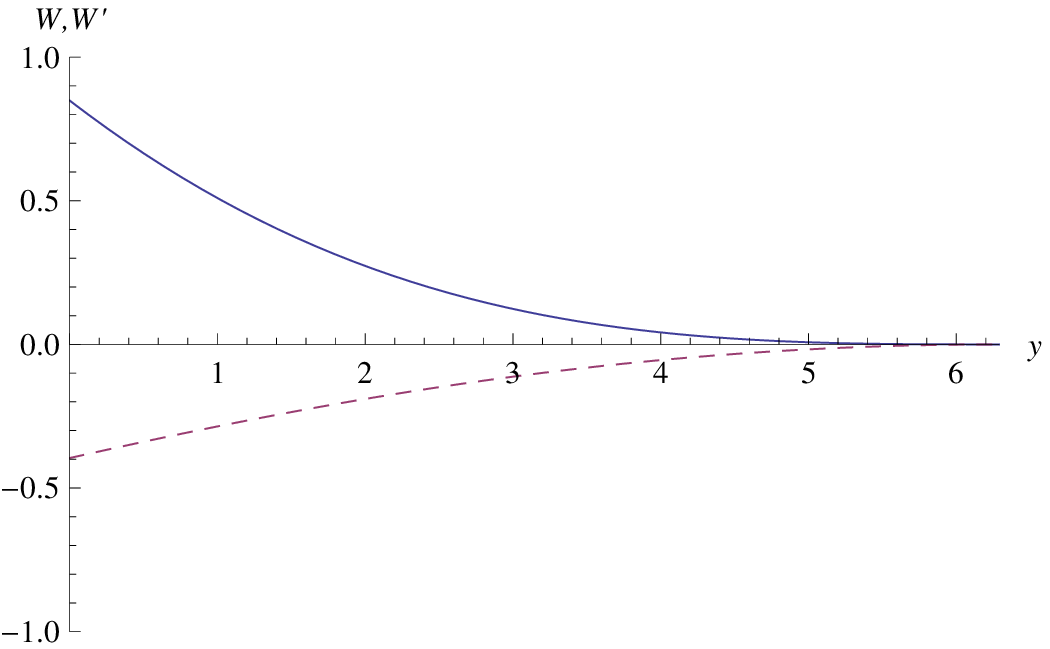}
  \caption{Solutions for the superpotential of the BPS system for $\mu^2=0.1$ and $g=0.1$ ($W' \ldots $dashed line).}
  \label{BPS_Sol}
 \end{center}
\end{figure}

As a last step, we want to solve directly the two BPS equations together with the superfield equation, both as a consistency check of our numerical calculation and as a demonstration of the BPS nature of the soliton solutions. We transform the superfield equation into a first order equation in the base space variable $y$ by multiplying it by $h_y^2$ and by using the chain rule $W_y = W_h h_y$ and arrive at the system of three first order equations 
\begin{equation} \label{BPS1a}
n h_y^2 (1+a) + \frac{1}{4} W_y = 0,
\end{equation}
\begin{equation} \label{BPS2a}
n a_y + g^2 \lambda^2 W = 0,
\end{equation}
(the two BPS equations), and
\begin{equation}\label{BPS3a}
\frac{1}{4} W_y^2 + g^2 \lambda^2 h_y^2 W^2 - \frac{4 \mu^2}{\lambda^2} h_y^2 h = 0
\end{equation}
(the superfield equation in base space), 
 where $y=r^2/2$. The boundary conditions we have to impose at the compacton boundary are
\begin{equation}
W(y_0)=0\; ,  \qquad  h(y_0) = 0
\end{equation}
whereas 
\be
W_y (y_0) =0\;  , \qquad h_y (y_0)=0 \; , \qquad a_y (y_0)=0
\ee
are then consequences of the above equations. We want to solve these equations via shooting from the compacton boundary.
Doing the power series expansion of the functions about the boundary $y_0$ and imposing the corresponding boundary conditions, we find
\begin{equation}
 h = \frac{\mu^2}{4 n^2 \lambda^2 (1+b_0)^2} (y-y_0)^2 + ... 
\end{equation}
\begin{equation}
 a = b_0 + \frac{g^2 \mu^4}{12 n^4 \lambda^2 (1+b_0)^3} (y-y_0)^4 + ... 
\end{equation}
\begin{equation}
W = -\frac{\mu^4}{3 n^3 \lambda^4 (1+b_0)^3} (y-y_0)^3 + ... 
\end{equation}
We conclude that the expansions for $h$ and $a$ are exactly the same as above. This demonstrates that, in this case, we will indeed get exactly the same soliton solutions as above. The two free constants $b_0$ and $y_0$ are again used to implement the two remaining boundary conditions $h(0)=1$ and $a(0)=0$ at the center. 
As a final check, we solve the system of equations (\ref{BPS1a}), (\ref{BPS2a}) and (\ref{BPS3a}) numerically via shooting from the boundary. In Figure (\ref{BPS_Sol}) we present the solution for the superpotential for the case $\mu^2=0.1$ with $g=0.1$ (the graphs of $h$, $a$ and the energy density are exactly like in Figure \ref{g01}). The values of the constants we get for this solutions are basically the same as before. We find a similar situation for other values of $\mu^2$ and (low) $g$. As explained before, we cannot use shooting from the boundary for high $g$ because of the problems with the expansion at the boundary.

\subsection{The potential $V=4h^2$}
\begin{figure}[h]
 \begin{center}
  \subfloat[Function $h$ and its derivative (dashed line).]{\includegraphics[width=0.6\textwidth]{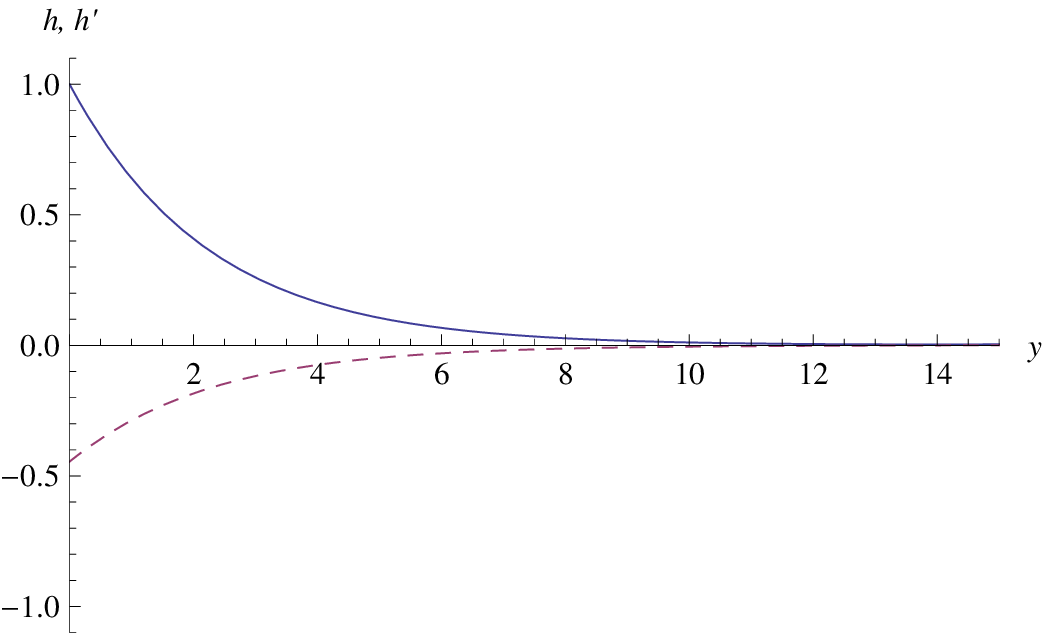}} \\
  \subfloat[Function $a$ and its derivative (dashed line).]{\includegraphics[width=0.6\textwidth]{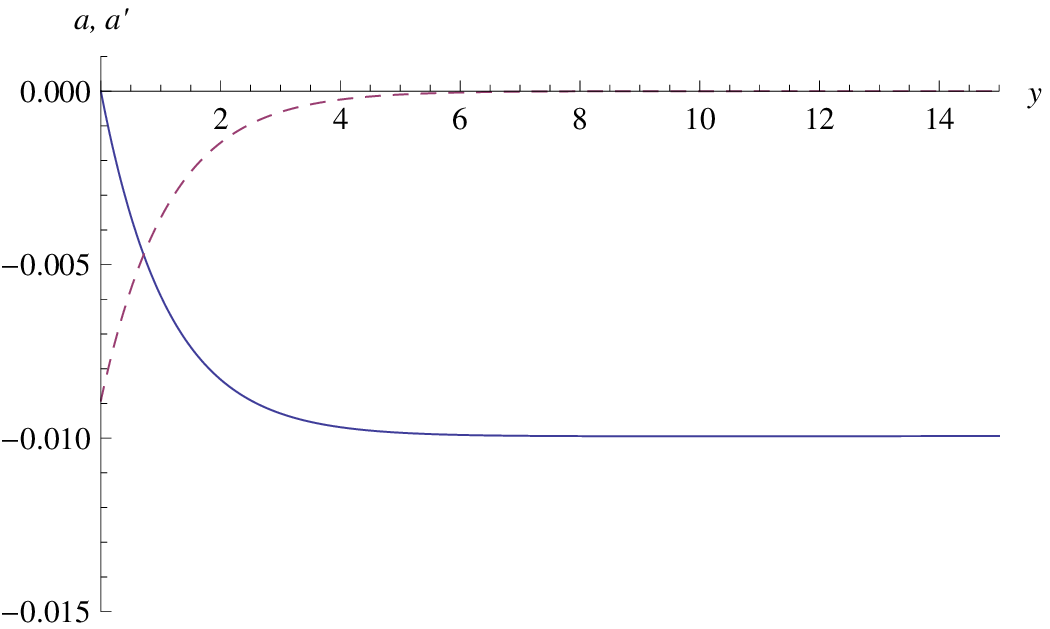}} \\
  \subfloat[Energy density.]{\includegraphics[width=0.6\textwidth]{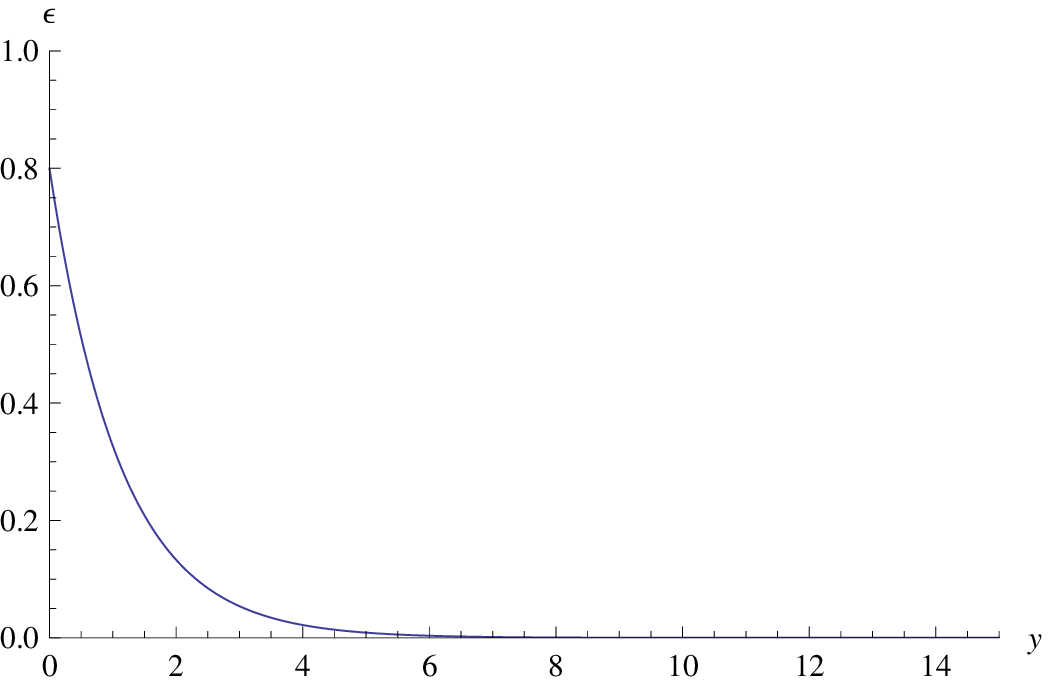}} \\
  \caption{Solutions for $\mu^2=0.1$ and $g=0.1$.}
  \label{Soln_01}
 \end{center}
\end{figure}

\begin{figure}[h]
 \begin{center}
  \subfloat[Function $h$ and its derivative (dashed line).]{\includegraphics[width=0.6\textwidth]{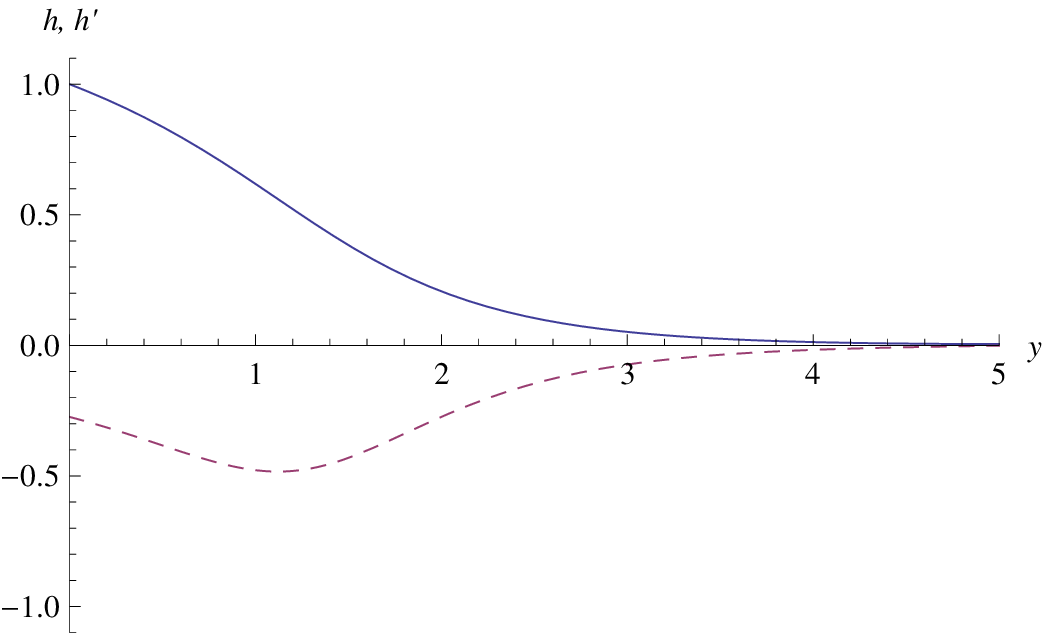}} \\
  \subfloat[Function $a$ and its derivative (dashed line).]{\includegraphics[width=0.6\textwidth]{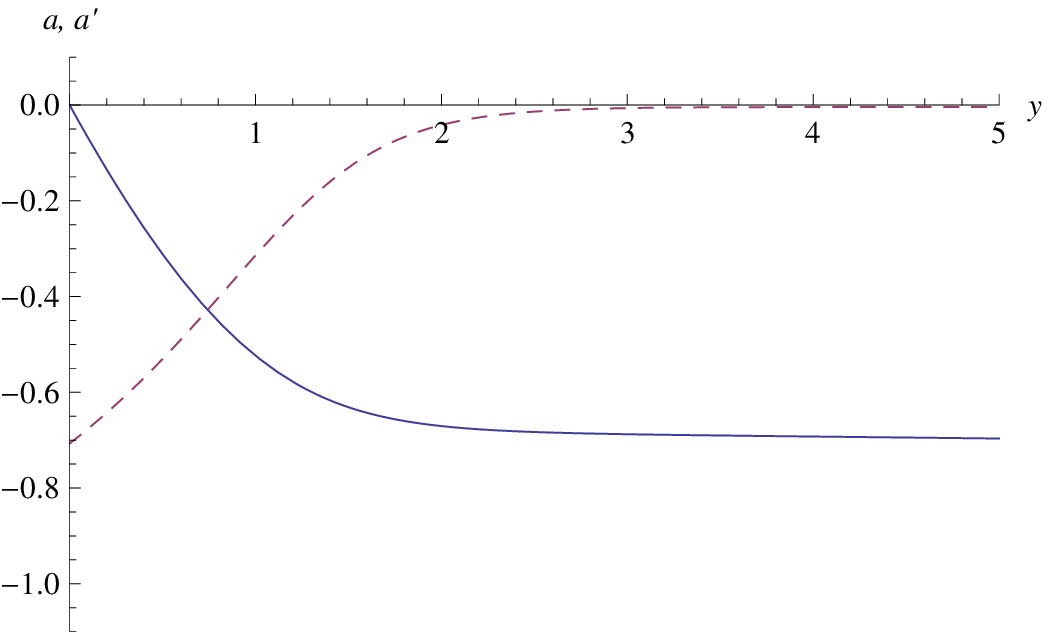}} \\
  \subfloat[Energy density.]{\includegraphics[width=0.6\textwidth]{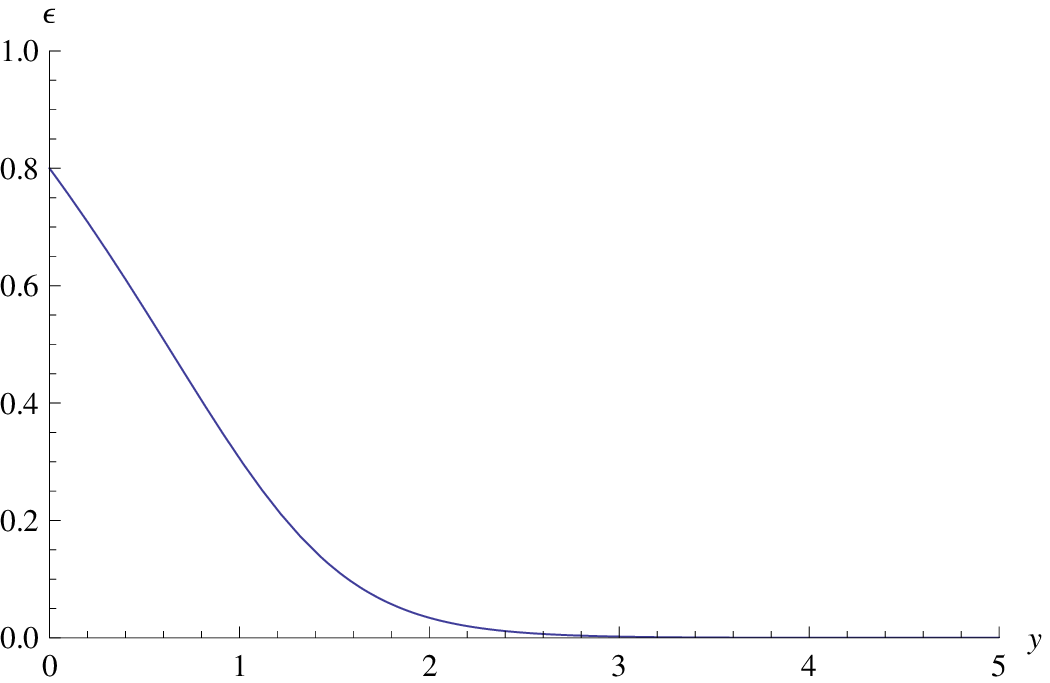}} \\
  \caption{Solutions for $\mu^2=0.1$ and $g=1$.}
  \label{Soln_1}
 \end{center}
\end{figure}

\begin{figure}[h]
 \begin{center}
  \subfloat[Function $h$ and its derivative (dashed line).]{\includegraphics[width=0.6\textwidth]{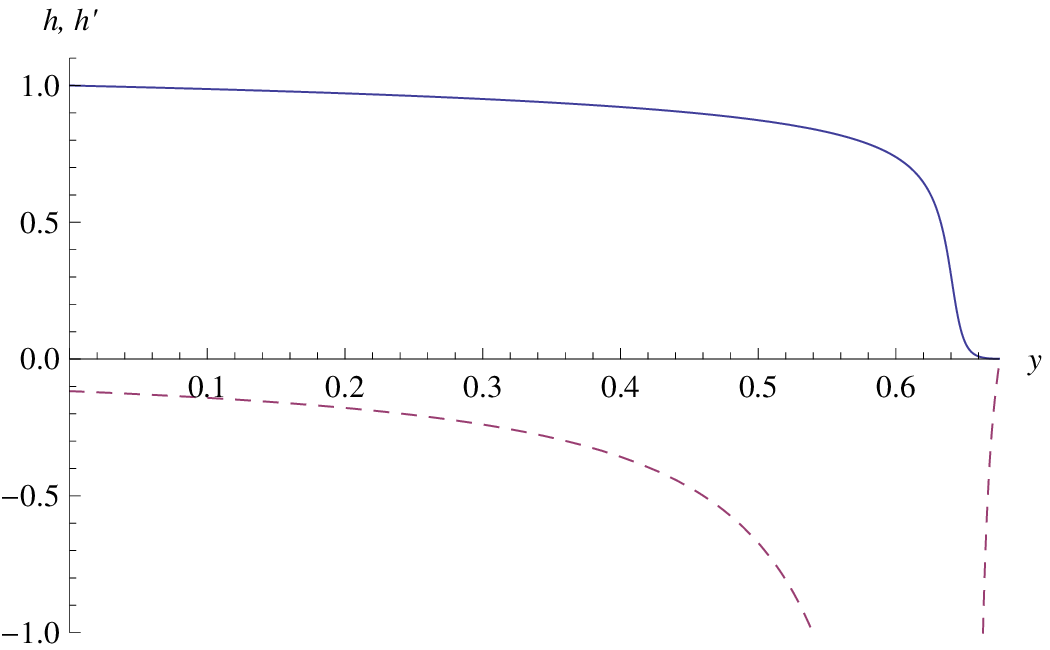}} \\
  \subfloat[Function $a$ and its derivative (dashed line).]{\includegraphics[width=0.6\textwidth]{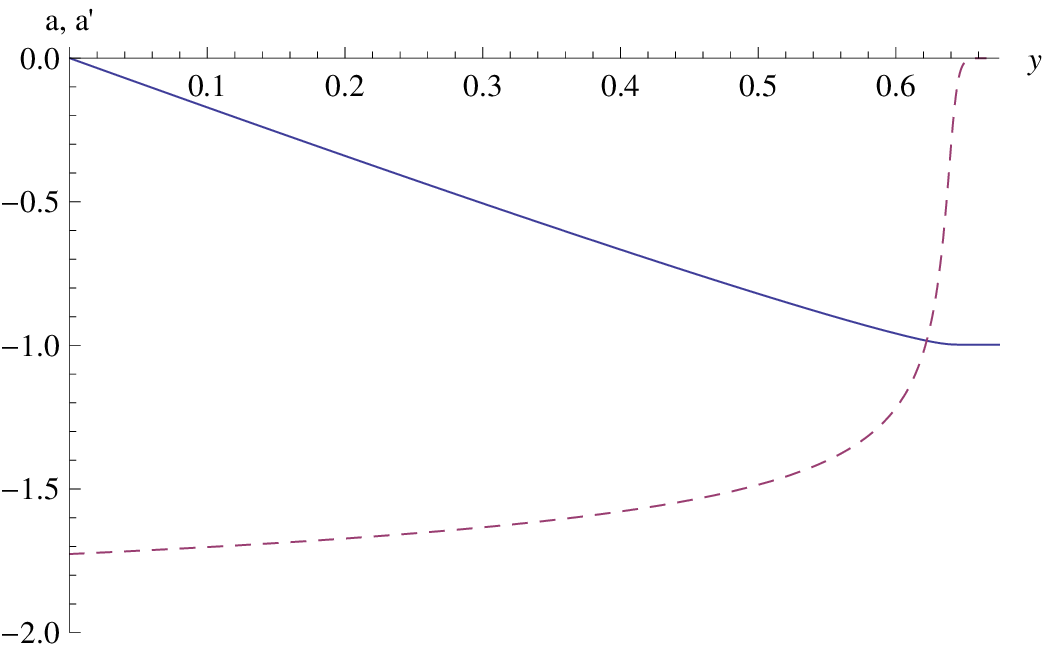}} \\
  \subfloat[Energy density.]{\includegraphics[width=0.6\textwidth]{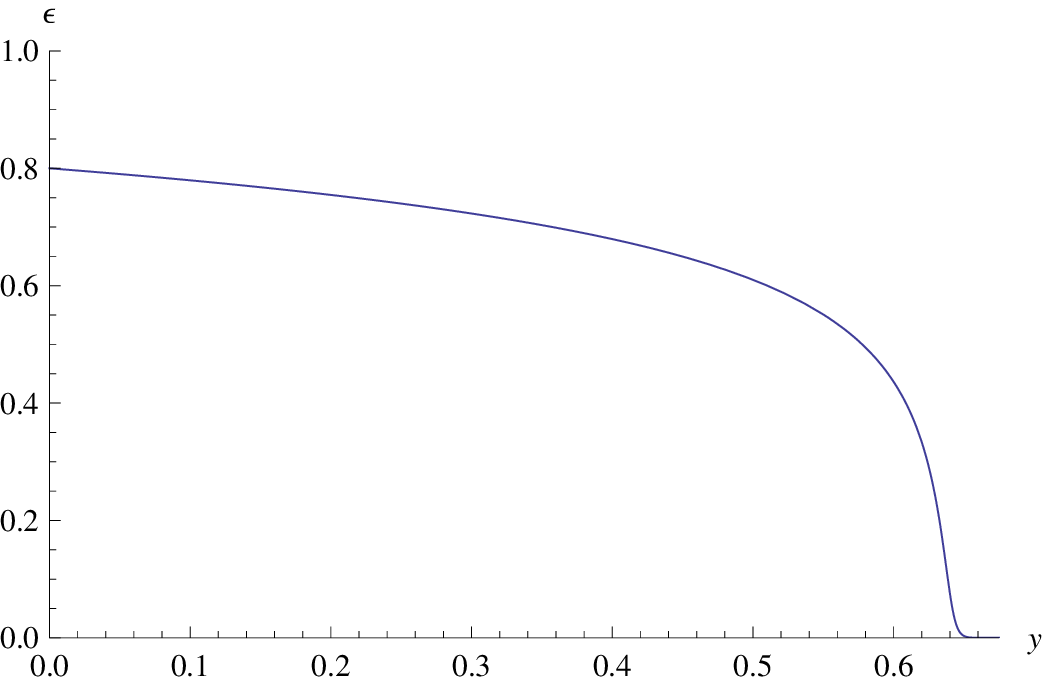}} \\
  \caption{Solutions for $\mu^2=0.1$ and $g=2$.}
  \label{Soln_2}
 \end{center}
\end{figure}

Here we want to briefly describe the numerical solution for the potential
\begin{equation}
V=(1 - \vec n \cdot \vec \phi)^2 = 4 h^2 .
\end{equation}
The system of static field equations for the spherically symmetric ansatz is
\begin{equation}
\partial_y [h_y (1+a)^2] - \frac{2 \mu^2}{n^2 \lambda^2} h = 0,
\end{equation}
\begin{equation}
a_{yy} = 4 \lambda^2 g^2 (1+a)h_y^2
\end{equation}
and the energy density reads
\begin{equation}
 \epsilon=2 \lambda^2 n^2 (1+a)^2 h_y^2 + 4 \mu^2 h^2 + \frac{1}{2g^2} n^2 a_y^2.
\end{equation}
In this case the potential is quartic in small fluctuations about the vacuum, therefore we expect exponential-type solutions, so we have to shoot from the center. Performing an expansion about the center and imposing
\begin{equation}
h(0)= 1, \qquad \qquad a(0)=0,
\end{equation}
we get
\begin{equation}
h(y) \sim 1 + f_1 y + \left( \frac{\mu^2}{n^2 \lambda^2} - b_1 f_1 \right) y^2,
\end{equation}
\begin{equation}
a(y) \sim b_1 y + 2 g^2 \lambda^2 f_1^2 y^2.
\end{equation}
Now we have the two free parameters $f_1$ and $b_1$ to satisfy the two boundary conditions in the limit $y\to \infty$. For the numerical calculation we choose $n=1$, $\lambda =1$, $\mu^2 = 0.1$ and $g=0.1,1,2$. Then we find soliton solutions for the values shown in table \ref{table4}.
\begin{table}
\begin{center}
 \begin{tabular}{|c|c|c|c|c|c|}
 \hline
  {\it g} & $f_1$ & $b_1$ & $E$ & $a(\infty )$ & Figure \\
\hline
 0.1 & - 0.44498 &  - 0.00892 & 5.60583 & - 0.009940  & \ref{Soln_01} \\
\hline
1 & -0.27393 &  -0.707 & 4.43440 & - 0.696595 & \ref{Soln_1} \\
\hline
2 &- 0.117166 & - 1.72637 & 2.71178 & - 0.99763 & \ref{Soln_2} \\
  \hline
\end{tabular}
\end{center}
\caption{Solutions of field equations for $\mu^2=0.1$ and $g=0.1,1,2$.}
\label{table4}
\end{table}
On the other hand, 
solving the superpotential equation (which has a global solution in this case) numerically, we find the BPS bound
$
E_B=2 \pi  |W(h=1)|
$
which agrees with the soliton energies $E$ within the shown precision in all three cases, therefore we do not show them separately. Apparently, the numerical convergence is even better for this potential. The result of the numerical integration of the soliton solutions is shown in Figures \ref{Soln_01}, \ref{Soln_1}, \ref{Soln_2}. For $g=0.1$ and $g=1$ the exponential approach to the vacuum is clearly visible. For $g=2$, on the other hand, the approach to the vacuum seems to be more like a compacton, and one wonders whether the exponential approach continues to hold for large $g$. We shall find in the next Section that the approach is, in fact, exponential, and the compacton-like appearance is due to a very fast exponential decay for large $g$, essentially like $\exp (-e^{g^2\lambda^2 }y)$.

Finally, we briefly comment on the magnetic flux and magnetic field for this case. Like in the case of the old baby Skyrme potential (Section V.A) and in Ref. \cite{GPS}, the magnetic flux $\Phi = 2\pi n a_\infty$ grows like $g^2$ in absolute value for small $g$ and approaches the constant value $-2\pi n$  in the limit of large $g$, in accordance with the exact results of Section IV.C.6. Further, the magnetic field itself in the limit of large $g$ behaves like the one in Section V.A, i.e., it is almost constant in the core of the soliton and drops to zero quickly in a thin shell of  rapid exponential decay, see Fig. \ref{Soln_2}.

\section{Exact solutions}

Here we want to show that for some fine-tuned choices of potentials there exist exact solutions both for the superpotential equation and for the corresponding BPS equations. It is in fact easy to find exact solutions to the superpotential equation. We just start with a given superpotential and calculate the potential from the superpotential equation (\ref{AV-eq2}). The disadvantage of this method is that it somewhat obscures the physical meaning of the coupling constants, because the potential now depends on them. For some fixed, given values of these constants, on the other hand, this is a perfectly valid method to produce exact solutions. Let us choose as a first example
\begin{equation} \label{W-exact}
W=\frac{1}{\lambda^2} h^b
\end{equation}
where $b>1$, then the potential is
\begin{equation}
V(h)=\frac{1}{2\mu^2\lambda^2}h^{2(b -1)} \left(\frac{b^2}{4} +g^2\lambda^2 h^2 \right) .\label{exact-pot}
\end{equation}
Besides being interesting in its own respect, this exact solution may help us to better understand the numerical solutions for potentials of the type $V \sim h^{2\alpha}$ studied in the previous Section, for the following reason. For small $g$, the first, $g$ independent part of the potential (\ref{exact-pot}) dominates and, therefore, the solution $W$ of the superpotential equation for the potential $V \sim h^{2\alpha}$ may be approximated by the exact solution (\ref{W-exact}). If we are interested only in the asymptotic behaviour, then the restriction to small $g$ can be lifted, because sufficiently close to the vacuum (i.e., for sufficiently small $h$), the first term (proportional to $b^2$) in the potential (\ref{exact-pot}) dominates even for large $g$. We shall see below that for the potential (\ref{exact-pot}) not only the superpotential is known, but also the spherically symmetric BPS equations can be solved exactly. We can, therefore, use the asymptotics of these exact solutions to determine exactly the asymptotics of solitons for the potentials $V \sim h^{2\alpha}$, and especially for the potential of Section V.C.

Before doing so, let us briefly mention that by this method we may also find global solutions for the superpotential equation for two-vacua potentials. Indeed, choosing, e.g., the superpotential
\begin{equation}
W=\frac{1}{\lambda^2} h^2(1-h)^2
\end{equation}
we find the potential
\begin{equation}
V=\frac{1}{2\mu^2\lambda^2} h^2(1-h)^2 \left[ (1-2h)^2 + g^2 \lambda^2 h^2(1-h)^2\right]
\end{equation}
where, as explained in Section IV.C.2, the extrema of $V$ and $W$ coincide. It follows from the results of Section IV.C.4 that, still, BPS soliton solutions do not exist for this potential. Besides, the existence of a global solution $W$ for this two-vacua potential $V$ is the consequence of a fine-tuning of the parameters of $V$. 

\subsection{Exact BPS solutions}
We shall find several examples where for an exact expression for the superpotential the spherically symmetric BPS equations
\begin{equation}
2nh_y(1+a)=-\frac{1}{2}W_h
\end{equation}
\begin{equation}
na_y=-g^2\lambda^2W
\end{equation}
have exact solutions, too. 

\subsubsection{A first example}
As a first example, we choose the superpotential $W=\frac{1}{\lambda^2}h^2$, with BPS equations
\begin{equation}
2nh_y(1+a)=-\frac{1}{\lambda^2} h 
\end{equation}
\begin{equation}
na_y=-g^2h^2 .
\end{equation}
Introducing $p=h^2$ we get
\begin{equation}
np_y(1+a)=-\frac{p}{\lambda^2} 
\end{equation}
\begin{equation}
na_y=-g^2p \label{ay} .
\end{equation}
Inserting the second formula into the first one gives
\begin{equation}
p_y(1+a)=\frac{a_y}{g^2\lambda^2} \;\; \Rightarrow \;\; p_y=\frac{1}{g^2\lambda^2} \frac{a_y}{1+a}=\frac{1}{g^2\lambda^2} \partial_y \ln (1+a) 
\end{equation}
and from (\ref{ay})
\begin{equation}
p_y = - \frac{n}{g^2} a_{yy}
\end{equation}
Then,
\begin{equation}
-na_{yy} = \frac{1}{\lambda^2} \partial_y \ln (1+a) \;\; \Rightarrow \;\; -n\partial_{yy} (1+a) = \frac{1}{\lambda^2} \partial_y \ln (1+a) 
\end{equation}
which leads to the first order equation
\begin{equation}
 \partial_y (1+a) =- \frac{1}{n\lambda^2} \ln [C(1+a)] . \label{sol ay}
\end{equation}
It can be further integrated 
\begin{equation}
\frac{1}{C}\int \frac{d(C(1+a)}{\ln[C(1+a)]}=-\frac{1}{n\lambda^2}(y-B)
\end{equation}
where $B$ and $C$ are integration constants. Thus finally 
\begin{equation}
\frac{1}{C} \mbox{Li} \;  [C(1+a(y))]  = -\frac{1}{n\lambda^2}(y-B) \label{sol a}
\end{equation}
where Li is the logarithmic integral function. The general solution for the profile function $h$ may be derived from (\ref{ay}) and (\ref{sol ay})
\begin{equation} \label{sol h2}
h^2=\frac{1}{g^2\lambda^2} \ln [C(1+a)] .
\end{equation}
It remains to determine the integration constants from the boundary conditions for a soliton solution. From $h(0)=1$, $a(0)=0$, and Eq. (\ref{sol h2}) we find
\be \label{C-val}
1 = \frac{\ln C}{g^2 \lambda^2} \quad \Rightarrow \quad C = e^{g^2 \lambda^2}
\ee
whereas from $a(0)=0$ and Eq. (\ref{sol a}) we get
\be
\frac{1}{C} \mbox{Li} \;  C  = \frac{B}{n\lambda^2} \quad \Rightarrow \quad B = n\lambda^2 e^{-g^2\lambda^2} \mbox{Li} \; (e^{g^2\lambda^2} ).
\ee
In order to find whether the fields take their vacuum values at a finite or infinite $y_0$, we first insert the boundary condition $h(y_0)=0$ into Eq. (\ref{sol h2}) to conclude
\be
C(1+a(y_0))=1
\ee
 and then use this result in Eq. (\ref{sol a}) to find
\be
\mbox{Li} \;  [C(1+a(y))] = \mbox{Li} \;  [1] = -\frac{C}{n\lambda^2}(y_0 -B) .
\ee
It is one of the properties of the logarithmic integral that $\mbox{Li}\, [1]=-\infty$, therefore we conclude that $y_0=\infty$. We may also determine the asymptotic value of $a(y)$,
\be
a(y=\infty) \equiv a_\infty = -1+C^{-1} = -1 + e^{-g^2 \lambda^2}
\ee
which implies $a_\infty \sim - g^2 \lambda^2 + {\cal O}(g^4 \lambda^4 )$ for small $g$ and $a_\infty \sim -1$ for large $g$, exactly reproducing the numerical findings of Section V.C (we remind the reader that the asymptotic behaviours of the exact solution of this section and of the numerical one of Section V.C are the same). 

To summarize, the soliton solution reads
\begin{equation}
a(y)=-1+\frac{1}{C} \mbox{Li}^{-1} \left( \mbox{Li}\; C - \frac{C}{n\lambda^2}y \right)
\end{equation}
\begin{equation}
h^2(y) =\frac{1}{g^2\lambda^2} \ln \left[ \mbox{Li}^{-1} \left( \mbox{Li}\; C - \frac{C}{n\lambda^2}y \right)\right]
\end{equation}
with $C$ given in (\ref{C-val}). The approach to the vacuum may be determined from the asymptotic behaviour of the logarithmic integral near one,
$$ \mbox{Li} (x) \approx \ln |1-x|, \;\;\; x \rightarrow 1 .$$
Hence, at $y \rightarrow \infty$ we have that $C(1+a(y)) \rightarrow 1$ and from (\ref{sol a})
\begin{equation}
\frac{1}{C} \ln |1-C(1+a(y))|= -\frac{y}{n\lambda^2}, \;\;\; y \rightarrow \infty
\end{equation}
Thus,
\begin{equation}
a(y) \approx \left( -1 +\frac{1}{C} \right)+ \frac{1}{C} e^{-\frac{Cy}{n\lambda^2}} \equiv a_{\infty}+ e^{-g^2 \lambda^2 } 
\exp \left( -\frac{e^{g^2 \lambda^2}y}{n\lambda^2} \right) ,
\end{equation}
\begin{equation}
h(y) \approx \frac{1}{g\lambda} \exp \left( -\frac{e^{g^2 \lambda^2}y}{2n\lambda^2}\right) ,
\end{equation}
where the last result comes from the second BPS equation
$na_y=-g^2h^2$.
It follows that these solitons (as well as the ones of Section V.C) are
exponentially localized, as was announced already in Section V.C.

\subsubsection{More examples of BPS solutions}

The BPS equations can, in fact,  be solved for all potentials (\ref{exact-pot}), i.e., for all superpotentials (\ref{W-exact}). The BPS equations are
\begin{equation}
2nh_y(1+a)=\mp \frac{b}{2\lambda^2} h^{b-1} 
\end{equation}
\begin{equation}
na_y=\mp g^2h^b
\end{equation}
where, depending on the particular case, one has to take the plus or minus sign. 
Repeating the previous computations we get
\begin{equation}
h^2=\frac{b}{2} \frac{1}{g^2\lambda^2} \ln [C(1+a)]
\end{equation}
\begin{equation}
\frac{1}{C}\int \frac{d(C(1+a))}{(\ln[C(1+a)])^{b/2}}=\mp \frac{g^2}{n} \left( \frac{b}{2 g^2\lambda^2} \right)^{b/2}(y-B)
\end{equation}
The boundary conditions at $y=0$ give
\begin{equation}
\ln C = \frac{2}{b} \lambda^2 g^2 .
\end{equation}
The boundary condition at $y=y_0$ gives again the asymptotic value of the magnetic function $a(y_0)$
\begin{equation}
C(1+a(y_0))=1 \;\; \Rightarrow \;\; a(y_0)=e^{-\frac{2}{b} \lambda^2 g^2} -1
\end{equation}
In order to calculate $y_0$ and $B$ one has to compute the integral, which strongly depends on the parameter $b$. For example for $b=4$ we get (we take the plus sign)
\begin{equation}
\mbox{Li} \; [C(1+a)] - \frac{C(1+a)}{\ln [C(1+a)]} = \frac{g^2}{n} \left( \frac{4}{2 g^2\lambda^2} \right)^{2}(y-B)
\end{equation}
The l.h.s. function $\mbox{Li} (x) - \frac{x}{\ln x}$ is a function which starts at zero value at the origin and then grows to infinity at $x=1$. Hence, $y_0=\infty$ i.e., it is a non-compacton configuration. Finally, as $a(0)=0$ we get
\begin{equation}
\mbox{Li} \; C - \frac{C}{\ln C} = - \frac{1}{n} \left( \frac{2}{ g\lambda^2} \right)^{2}B
\end{equation}
or 
\begin{equation}
B= - n \left( \frac{ g\lambda^2}{2} \right)^{2} \left( \mbox{Li} \; e^{\frac{1}{2} \lambda^2 g^2} - \frac{2e^{\frac{1}{2} \lambda^2 g^2}}{\lambda^2g^2} \right) .
\end{equation}

 \section{Conclusions}
In this article we investigated in depth the gauged version of the BPS baby Skyrme model, i.e., the BPS baby Skyrme--Maxwell system. We found that, like the ungauged model, the gauged model still has infinitely many symmetries, both Noether and non-Noether ones. Further, it continues to have a BPS bound and to support BPS soliton solutions saturating this bound. The BPS bound is, however, quite different from other, known BPS bounds. Known BPS bounds either bound the energy (or euclidean action) in terms of the topological degree only (e.g., instantons, O(3) nonlinear sigma model) or involve both the topological charge density and the potential of the theory (e.g., scalar field domain walls, or the ungauged BPS baby Skyrme model). Here, on the other hand, we find a BPS bound in terms of the topological charge density and an auxiliary function which is related to the potential via a first order differential equation. We called the auxiliary function "superpotential" and its defining differential equation "superpotential equation" due to their similarity with the corresponding equation in supergravity, as explained Section IV.A. A further consequence of this new type of BPS bound is that it is a rather nontrivial problem whether for a given potential BPS soliton solutions exist, at all. One necessary condition is the existence of the superpotential on the whole target space (i.e., a global solution of the superpotential equation (\ref{AV-eq2})), but we found examples where BPS soliton solutions do not exist despite a globally existing superpotential, so this condition is, in general, not sufficient. Specifically, we found that for potentials with more than one vacuum BPS solitons never exist. The ungauged model with multi-vacua potentials, on the other hand, supports both a BPS bound and BPS solitons, so we find the interesting result that for systems where the matter sector does not have a unique vacuum, the inclusion of the electromagnetic interaction destabilizes the "particles" (topological solitons) of the theory. For models with monotonously growing potentials, instead, we found both a nontrivial BPS bound and BPS soliton solutions in all cases we considered, motivating the conjecture that this will always be true. 

We studied rotationally symmetric solitons numerically for some concrete potentials and  found in all cases that the soliton energies, indeed, saturate the corresponding BPS bound. We also found that the soliton energies decrease with increasing electromagnetic coupling (e.g., like $g^{-1}$ for the old baby Skyrme potential, see Section V.A). If the superpotential is known exactly, we found that, at least in some cases, the whole system of BPS equations can then be solved exactly, providing us with analytic, explicit expressions for the solitons, see Section VI. A further quantity which can be calculated exactly once the superpotential is known is the magnetic flux, see Section IV.C.6. Finally, we want to point out that the BPS bounds we found for the gauged BPS baby Skyrme models for different potentials provide, at the same time, BPS bounds for the corresponding full gauged baby Skyrme models which will, in many occasions, provide tighter bounds that the bounds already known (see our discussion in Section II.B).

The model has shown a surprisingly rich mathematical structure and there remain many interesting open problems which deserve further investigation. One first, obvious issue is to better understand under which conditions BPS soliton solutions will exist. That is to say, to find rigorous mathematical answers to the following two questions,
\\
$i)$ for which potentials $V(h)$ the superpotential equation (\ref{AV-eq2}) has global solutions in the whole interval $h \in [0,1]$ (i.e., on the whole target space), and
\\
$ii)$ which additional conditions the potentials have to obey such that BPS soliton solutions exist. 

Another question of interest is whether the BPS property of the model may be related to some further structure as, e.g., supersymmetry, as is frequently the case. In this respect, already the ungauged model involves a surprise. Indeed, while it was found recently that the full baby Skyrme model allows for a supersymmetric extension \cite{susy-bS}, this supersymmetric extension is not possible for the restricted (i.e., BPS) baby Skyrme model, which indicates that the relation between BPS equations and supersymmetry is more involved for field theories with non-standard kinetic terms.

From a more physical point of view, the most important question is, of course, if and to which degree the structure found in this model can be generalized to the Skyrme model in 3+1 dimensions. A first important observation is that, as was already discussed in the introduction, the ungauged Skyrme model, too, has a submodel (the BPS Skyrme model) sharing all the nontrivial features with the BPS baby Skyrme model (integrability, BPS bound and exact BPS solutions). In this sense, the hope to be able to generalize some of the results of the present paper to 3+1 dimensions is well-founded. The gauged version of the BPS Skyrme model will, nevertheless, most likely present additional difficulties, where the most obvious one is related to the fact that the magnetic field is a pseudoscalar in 2+1 dimensions, whereas it is a pseudovector in 3+1 dimensions. A further question of interest will be whether the gauged model in 3+1 dimensions maintains all the symmetries of the ungauged model, as is the case in 2+1 dimensions. In any case, this problem shall be investigated in future publications, where we hope that the detailed analytical and numerical results developed in the present article should be instrumental in the understanding and investigation of the 3+1 dimensional system. 

Finally, another question of considerable interest is, in our opinion, whether BPS bounds of the type found in the present article can be applied in a more general context, i.e., outside the field of Skyrme type models.

{\em Note added:} While finishing this paper we became aware of the preprint \cite{step1}, where the BPS bound we derived in Section IV.A was found using completely different methods, extending previous results of the so-called Bogomolny decomposition for the ungauged model \cite{step2}.

\section*{Acknowledgement}
The authors acknowledge financial support from the Ministry of Education, Culture and Sports, Spain (grant FPA2008-01177), 
the Xunta de Galicia (grant INCITE09.296.035PR and
Conselleria de Educacion), the
Spanish Consolider-Ingenio 2010 Programme CPAN (CSD2007-00042), and FEDER. 
CN thanks the Spanish
Ministery of
Education, Culture and Sports for financial support (grant FPU AP2010-5772).
Further, AW was supported by polish NCN grant 2011/01/B/ST2/00464. Finally, the authors thank M. Speight for helpful comments and for pointing out an error in an earlier version of the manuscript.

\end{document}